\documentclass[aps,prb,reprint,showpacs,superscriptaddress,nofootinbib,floatfix,noshowpacs,showkeys,longbibliography]{revtex4-1}
\usepackage[colorlinks=true,linkcolor=blue,urlcolor=blue,citecolor=blue,pdfusetitle]{hyperref}
\usepackage[utf8]{inputenc}
\usepackage[english]{babel}
\usepackage{amsmath}
\usepackage[caption = false]{subfig}
\usepackage{graphicx,epstopdf}
\usepackage{blindtext}
\usepackage{lipsum}
\usepackage{amsfonts}
\usepackage{bbm}
\usepackage{amssymb}
\usepackage{enumerate}
\usepackage{color}
\usepackage{latexsym}
\usepackage{times,txfonts}
\usepackage{natbib}
\usepackage[usenames,dvipsnames]{xcolor}

\usepackage{tikz}

\begin{document}

\title{Doping Topological Dirac Semimetal with magnetic impurities:\\ 
electronic structure of Mn-doped Cd$_3$As$_2$}

\author{H. Ness}
\affiliation{Department of Physics, King's College London, Strand, London WC2R 2LS, UK}

\author{I. Leahy}
\affiliation{National Renewable Energy Laboratory, Golden, Colorado 80401, USA}

\author{A. Rice}
\affiliation{National Renewable Energy Laboratory, Golden, Colorado 80401, USA}

\author{D. Pashov}
\affiliation{Department of Physics, King's College London, Strand, London WC2R 2LS, UK}

\author{K. Alberi}
\affiliation{National Renewable Energy Laboratory, Golden, Colorado 80401, USA}

\author{M. van Schilfgaarde}
\affiliation{National Renewable Energy Laboratory, Golden, Colorado 80401, USA}

\begin{abstract}
{
The prospect of transforming a Dirac topological semimetal (TSM) into a Weyl TSM phase, 
following doping by magnetic impurities, is central to TSM applications.
The magnetic field from polarized $d$ levels of magnetic impurities produces a field with 
a sharp local structure.  
To what extent magnetic impurities act in the same manner as an applied field and what are
the effects of such a field on the electronic structure of a Dirac TSM
is the subject of this paper.
}
We present electronic structure calculations of bulk Cd$_3$As$_2$ with
substitutional doping of Mn impurities in the dilute alloy range.
Quasi-particle $GW$ (QS$GW$) \emph{ab-initio} electronic structure calculations are 
used
in conjunction with $k \cdot p$ model Hamiltonian calculations.
As expected, we observe the splitting of the Dirac points into pairs of Weyl points
following the doping with Mn.
We also show that the electronic structure of Mn-doped Cd$_3$As$_2$ can be emulated 
by the electronic structure of pristine Cd$_3$As$_2$ with an appropriate external 
magnetic field.
Some properties of the conductivity of 
bulk Cd$_3$As$_2$ for different magnetic field orientations are also investigated.
{
Our results inform future opportunities for unique 
device functionality} 
based on band structure tuning not found in conventional magnetic Weyl TSM.

\end{abstract}


\maketitle

\section{Introduction}
\label{sec:intro}

Three dimensional topological semimetals (TSMs) encompass a broad class of materials 
that are generally characterized by the presence of linear band touching nodes near their 
Fermi levels \cite{Armitage:2018}. 
These nodes are stabilized by the presence of specific symmetries. The exact details 
of their electronic structures, however, differ based on the combination of symmetries 
that are present, leading to several TSM sub-classes \cite{Armitage:2018,Li:2020}. 
While the sub-classes typically exhibit many similar properties (i.e., high carrier 
mobility or broadband absorption), differences can also support diverse phenomena 
that may be exploited for various applications \cite{Armitage:2018,Liang:2015,Wang:2020,He:2022,Bernevig:2022}.

An important question to ask is whether a TSM can be transformed from one sub-class to 
another by manipulating a specific symmetry. Such a capability can be advantageous for 
applications where it is desirable to add one or more properties. 
A prime example is the formation of a Weyl TSM from a Dirac TSM. Dirac TSMs exhibit 
both time reversal symmetry (TRS) and spatial inversion symmetry (IS) that help to 
support the presence of fourfold degenerate bands at the Dirac node(s) located along 
some high symmetry line(s) in reciprocal $k$-space. 
When either TRS or IS is broken, the Dirac nodes are instead represented as constituent 
Weyl nodes (with twofold band degeneracy and opposite chirality). 
Weyl TSMs exhibit several properties that are not present in Dirac TSMs, including robust 
Fermi arc surface states, the chiral anomaly, the anomalous Hall effect and 
the bulk photovoltaic effect \cite{Armitage:2018}.

{
An example of such a Dirac to Weyl TSM transformation was demonstrated 
in Bi-Sb compounds by Au ion implantation leading to IS breaking} \cite{Lee:2022}. 
{
Attempts to transform the Dirac TSM Cd$_3$As$_2$ into a Weyl TSM EuCd$_2$As$_2$,
with the subtitution of a magnetic element for Cd, have been made} \cite{Wang:2019,Ma:2019}.
{
However recent studies have shown that EuCd$_2$As$_2$ is a magnetic, small gap, semiconductor 
rather than a Weyl TSM} \cite{Santos:2023,Shi:2024,Nishihaya:2024}.

It is therefore interesting to know
if and how a Dirac TSM can 
be transformed into a Weyl TSM through the addition of magnetic impurities. 
In the dilute alloy limit, such impurities have the potential to introduce a large local 
magnetic field and break TRS without significantly altering other aspects of the material structure. 
{
In this work, we explore and clarify the full extent to which substitutional doping with 
magnetic impurities transforms Dirac nodes into Weyl nodes and the relationships between 
the orientation of the magnetic moment, the electronic structure, and other properties, 
such as conductivity. 
We use Mn impurities in bulk Cd$_3$As$_2$ as a prototypical three dimensional Dirac system
\cite{ARice:2025}.
Cd$_3$As$_2$ \cite{Liu:2014} contains two Dirac points on the $\Gamma Z$ line in the Brillouin zone,
protected by the $C_4$ symmetry about $z$.  
For an ideal, undoped lattice, the Fermi energy $E_F$ passes though the Dirac points. 
}

{
As Mn and Cd both have two valence electrons, Mn substituting for Cd does not electrically dope Cd$_3$As$_2$.  
However, Mn has a filled shell of 3d states in the majority channel, with the minority channel empty.  
Thus each Mn has a local magnetization of 5 $\mu_B$, similar to MnTe.  
The majority and minority levels are split 
{{by a few eV}} 
roughly evenly about the Fermi energy
according to the QS$GW$ approximation described below. 
Thus both states are well removed from $E_F$ and have little effect on states around
$E_F$, except for the large effective magnetic field from the magnetization of the Mn $d$ levels.  
The Mn local moment polarizes the Mn and As $sp$ levels, and we may therefore expect it to behave 
similarly to an external magnetic field, breaking the TRS.  
To what extent Mn acts in the same manner as an applied field is a primary subject of this paper.

As we show below, the effective field generated by local moment can be very large compared to external fields 
accessible in the laboratory.  
However the net contribution from Mn will depend on the doping, and to what extent the Mn spins are ordered.  
For the low doping regime we consider here the spins should be completely disordered; 
however, the spin susceptibility will be greatly enhanced by the partial alignment of the spins on application of a field.  
For the electronic structure we consider an idealized case: substituting a single Mn 
on a Cd site in the unit cell of Cd$_3$As$_2$ 
(i.e. doping $\sim$ 2 \%), 
aligned ferromagnetically.
}

We use \emph{ab-initio} electronic structure calculations combined with
model Hamiltonian calculations to understand the transformation of Dirac
points into Weyl points in bulk Cd$_3$As$_2$. We show that the electronic
structure of Mn-doped Cd$_3$As$_2$ can be well reproduced from the electronic
structure of pristine Cd$_3$As$_2$ with an appropriate external magnetic
field. 
We also briefly study some properties of the conductivity of bulk Cd$_3$As$_2$
for different orientations of an applied magnetic field.
 
The paper is organised as follows:
In Section \ref{sec:calc} we present and discussion the results of our calculations
for the electronic structure and the conductivity. Conclusions are presented
in Section \ref{sec:ccl}. Additional information are provided in the appendices, about:
technicality for the \emph{ab-initio} calculations in Appendix \ref{app:questaal},
the model Hamiltonian in Appendix \ref{app:kdotp} and 
linear response conductivity in Appendix \ref{app:sigma}.

\section{Calculations and discussion}
\label{sec:calc}

{In this section, we show how the electronic structure of pristine Cd$_3$As$_2$
is modified by the inclusion of Mn magnetic impurities, i.e. the splitting of the Dirac
points into pairs of Weyl points. We also study how an applied magnetic field can
have similar effects on the band structures. The calculations are performed by using
both an \emph{ab-initio} electronic structure method and model Hamiltonians.}

\subsection{\emph{Ab-initio} electronic structure}
\label{sec:estruc}

First principles electronic structure calculations have been performed using the Questaal package \cite{questaal:web}.
Questaal is an all-electron method, with an augmented wave basis consisting of partial waves inside augmentation
spheres based on the linear muffin-tin orbital technique \cite{Pashov:2020}.
{
Calculations have been done at different levels of density functional theory (DFT), i.e. LDA and GGA.
Going beyond DFT, we have
also performed Quasiparticle Self-consistent \emph{GW} (QS\emph{GW}) calculations \cite{Faleev:2004}.  
QS\emph{GW} provides an effective way to implement the \emph{GW} approximation without relying on a lower level theory 
as a starting point \cite{MvS:2006,Kotani:2007}.
}

We consider a primitive unit cell, for bulk Cd$_3$As$_2$, consisting of 80 atoms 
(48 atoms of Cd and 32 of As) as shown in panel (a) of Figure \ref{fig:qsgwbands}.
See Appendix \ref{app:questaal} for information about the unit
cell and $k$-space mesh.

\begin{figure}
\centering
\text{(a)}\includegraphics[width=32mm]{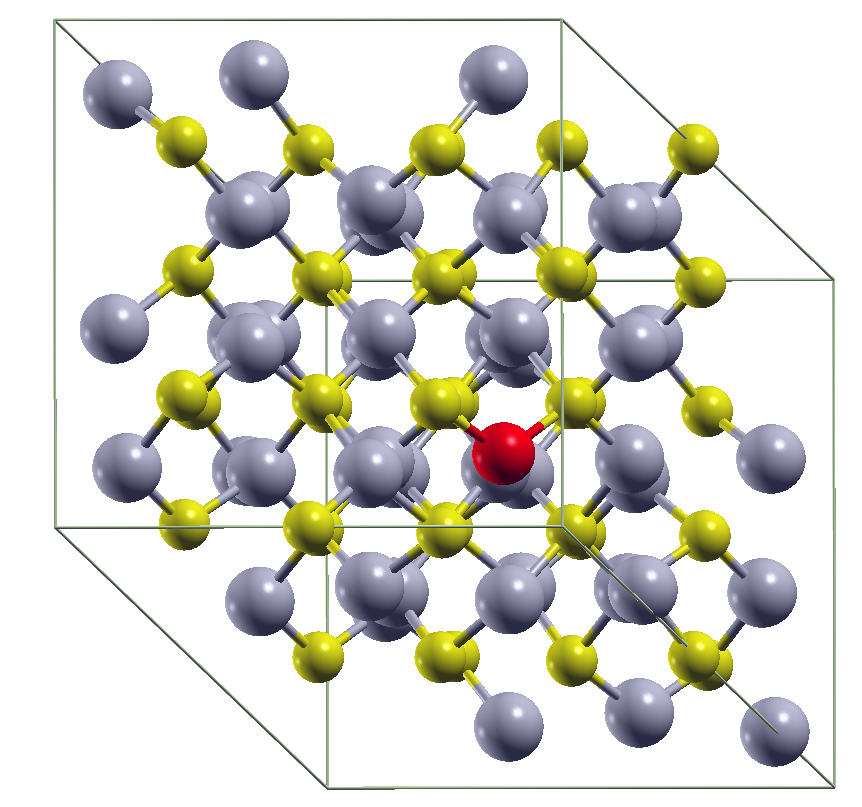}\includegraphics[width=37mm]{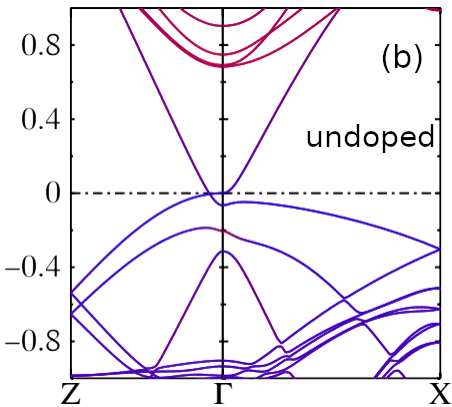}
\includegraphics[width=37mm]{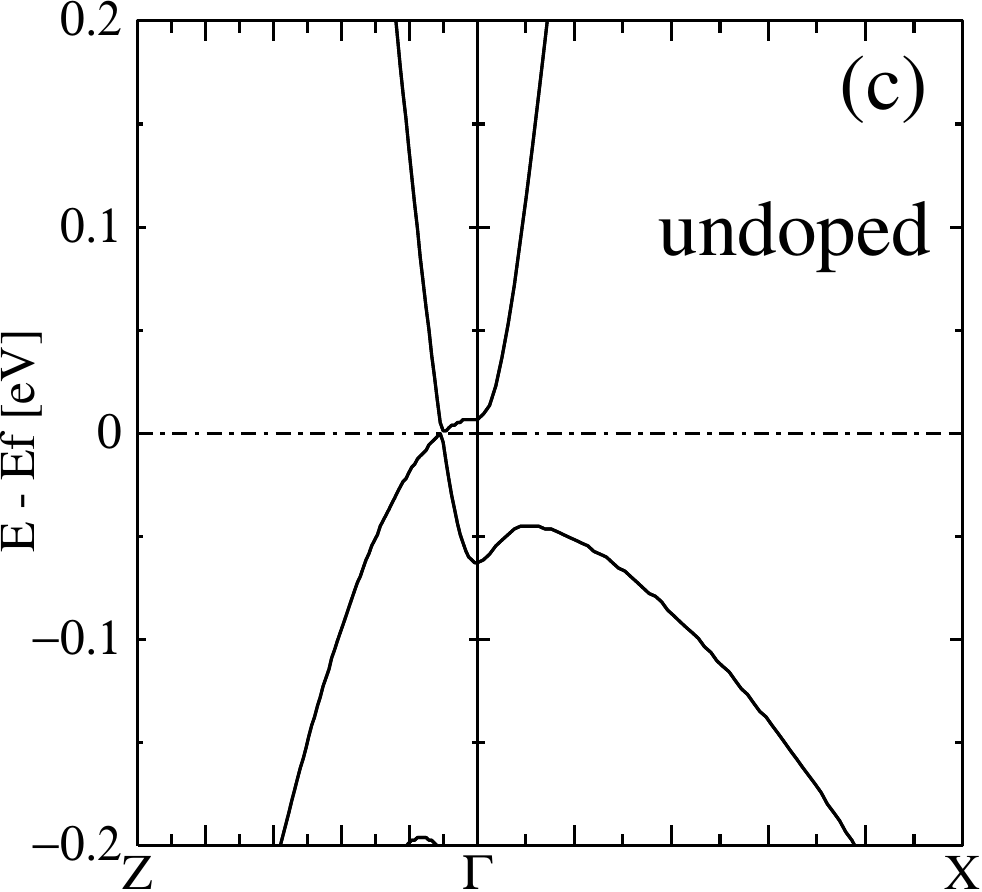}\includegraphics[width=39mm]{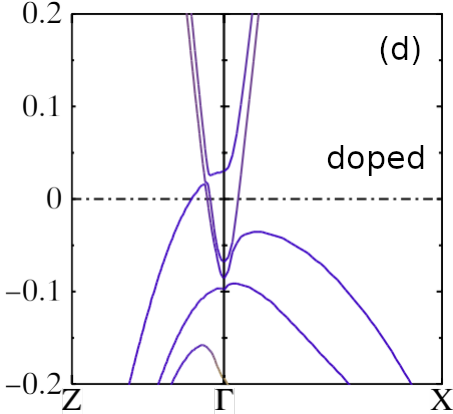}
\caption{\label{fig:qsgwbands}
Panel (a): 
Ball-and-stick representation of the supercell for Cd$_3$As$_2$.  Cd atoms are shown in grey, As atoms in yellow.
The red atom is Cd for the pristine structure, and is Mn for the 2 \% doped crystal. 
Panel (b):
QS$GW$ electronic band structure of Cd$_3$As$_2$ for high symmetry $k$-space path $Z \Gamma X$.
Energies, in eV, are measured from the Fermi level $E_F=0$ reference energy where the Dirac point lies. 
The two Dirac points for the entire pristine Cd$_3$As$_2$ are located in $k$-space at $(0,0,\pm k_D)$
with $k_D \sim 0.03\ \AA^{-1}$. Color weighting: red for Cd $5s$ orbital and blue for As $4p$.
Panel (c):
Zoom in energy around the Dirac point.
Panel (d): 
QS$GW$ band structure of Mn-doped Cd$_3$As$_2$ around the Dirac
point observed in the undoped case.
Bands are split by a Zeeman-like effect due to the magnetic Mn impurity and the Dirac point seems
to have disappeared.
}
\end{figure}

{Before exploring the effects of Mn doping, we verify the known Cd$_3$As$_2$ band structure using both DFT and QS$GW$.}
Our DFT calculations (not shown) agree with previously reported band-structures 
\cite{Wang:2013,Ali:2014,MoscaConte2017,Crassee:2018,Yue:2019,Kulatov:2021,Brooks:2023}.  
Figure \ref{fig:qsgwbands} shows our QS$GW$ calculated band structures.
Panels (b) and (c) in Fig.\ref{fig:qsgwbands} show the QS$GW$ bands
for undoped Cd$_3$As$_2$.
Two Dirac points (i.e. crossing of 4 degenerate bands of mostly $4p$ As character) are found at $E_F$ 
along 
the $Z \Gamma$ line, they are located at $k=(0,0,\pm k_D)$ 
{with $k_D \sim 0.029\ \AA^{-1}$, 
a slightly smaller value than those based on DFT calculations 
}
\cite{Wang:2013,Ali:2014,MoscaConte2017,Yue:2019}.  

{The band velocity around the $\Gamma$-point and around the Dirac points
are given in Table \ref{table:vF} for pristine Cd$_3$As$_2$.
The QS$GW$ calculations provide, for at least one branch of the Dirac crossing
bands, a velocity of $\sim 9-10 \times 10^5$ [m/s], in agreement with values
found in the literature \cite{Jeon:2014,Liang:2015,Nelson:2023}.
By contrast, the band velocity is severely underestimated in DFT.  
This is to be expected because of the local potential DFT generates. 
DFT underestimates bandgaps in semiconductors \cite{Maksimov:1989}
and the Fermi velocity at the Dirac point in graphene \cite{mvsgraph:2011}. 
A local LDA potential in the auxiliary DFT Hamiltonian results in underestimates of the bandgap, as  
clearly shown for a few semiconductors in Ref \cite{Gruning:2006}.

Shortcomings in the LDA and GGA are also reflected in the location of the Cd 5$s$ states: 
in QS\emph{GW} they are higher in energy than in DFT. 
When these states are too low, they may repel the As 4$p$ states and
contribute to the underestimate of the Fermi velocity.
Also admix too much of Cd $s$ into the states in the vicinity of the Dirac
point.
Furthermore, the band inversion mechanism leading to the crossing at $k_z=k_D$ along the $(0,0,\pm k_z)$ lines
involves a parabolic and an inverted parabolic bands with mostly As 4$p$ character.
The effective mass of electron-like band is smaller than the effective mass of hole-like
band, therefore around the Dirac point, one branch (annoted $Z\Gamma \text{ down}$ in Table \ref{table:vF})
of the crossing bands is associated with a larger velocity than the other 
branch (annoted $Z\Gamma \text{ up}$ in Table \ref{table:vF}).
}

We now turn to the case of Cd$_3$As$_2$ doped by magnetic impurities.
We consider that, for the Mn-doped system, one Cd atom in the unit cell is replaced by Mn, 
i.e. corresponding to a doping of 1/48 $\sim$ 2 \%.
The band structure of the Mn-doped Cd$_3$As$_2$ is substantially different from the undoped system, 
see panel (d) in Figure \ref{fig:qsgwbands}.  
Bands are split by a Zeeman-like effect owing to the polarization of the As 4$p$ states 
induced by the magnetic Mn impurity.

The Dirac point disappears from the symmetry protected $Z \Gamma$ $k$-space line.
Gapped bands occur around $E-E_F=25$ meV close to the $\Gamma$ point on the $Z \Gamma$ $k$-path.

{The majority(minority) Mn $d$ states consists of occupied (empty) states, centered at $E_F - 4.5$ eV
($E_F + 3.5$ eV) respectively.
They make the system ferromagnetic with a net moment of 5.0 $\mu_B$.  
The Mn local moment in the Mn augmentation sphere is found to be 4.4 $\mu_B$, the remaining 0.6 $\mu_B$ 
decaying away from the Mn.  
According to the Stoner picture, we should expect the $d^{\uparrow}-d^{\downarrow}$ energy splitting to be 
$I{\cdot}M$, where $I$ is the Stoner parameter and $M$ the local moment.  
As a rule of thumb, $I{=}1$ eV for 3$d$ transition metals and would lead to $d^{\uparrow}-d^{\downarrow}=8.8$ eV.  
Thus the Stoner splitting is qualitatively correct, though it overestimates slightly the actual splitting.  
Both $d^{\uparrow}$ and $d^{\downarrow}$ are very localized, each with a bandwidth of $\sim 2$ eV.  
The dispersion establishes that the Mn $d$ band hybridizes to some extent with the Cd and As (the Mn-Mn separation 
is 12.6 $\AA$, too large for direct coupling) and induces a small local moment on those sites.
Two of the four As nearest neighbors to Mn acquire a moment of about 0.025 $\mu_B$.  
For comparison, the Mn local moment in LDA is calculated to be 3.9 $\mu_B$ and the $d^{\uparrow}$ and $d^{\downarrow}$ 
band centers fall at approximately $E_F - 3.0$ eV and ($E_F + 1.0$ eV), respectively.  
The splitting is small enough that states near $E_{F}$ have non-negligible Mn $d$ character.  
This shows only that the LDA does not provide an adequate description of
the magnetic electronic structure.
}

As previously mentioned, the magnetic field coming from polarized Mn $d$ levels has a sharp local 
(in space) structure,
somewhat different from what is expected from a macroscopic applied field.
However, it is interesting to see to which extent this field acts in a similar manner as an applied 
external field.

In order to get more insight into the band structure of Mn-doped Cd$_3$As$_2$, 
we consider next how the band structure of pristine Cd$_3$As$_2$ is modified by the presence of an 
applied external magnetic field (the latter emulating the presence of the Mn magnetic impurity).

\begin{figure}
\centering
\includegraphics[width=80mm]{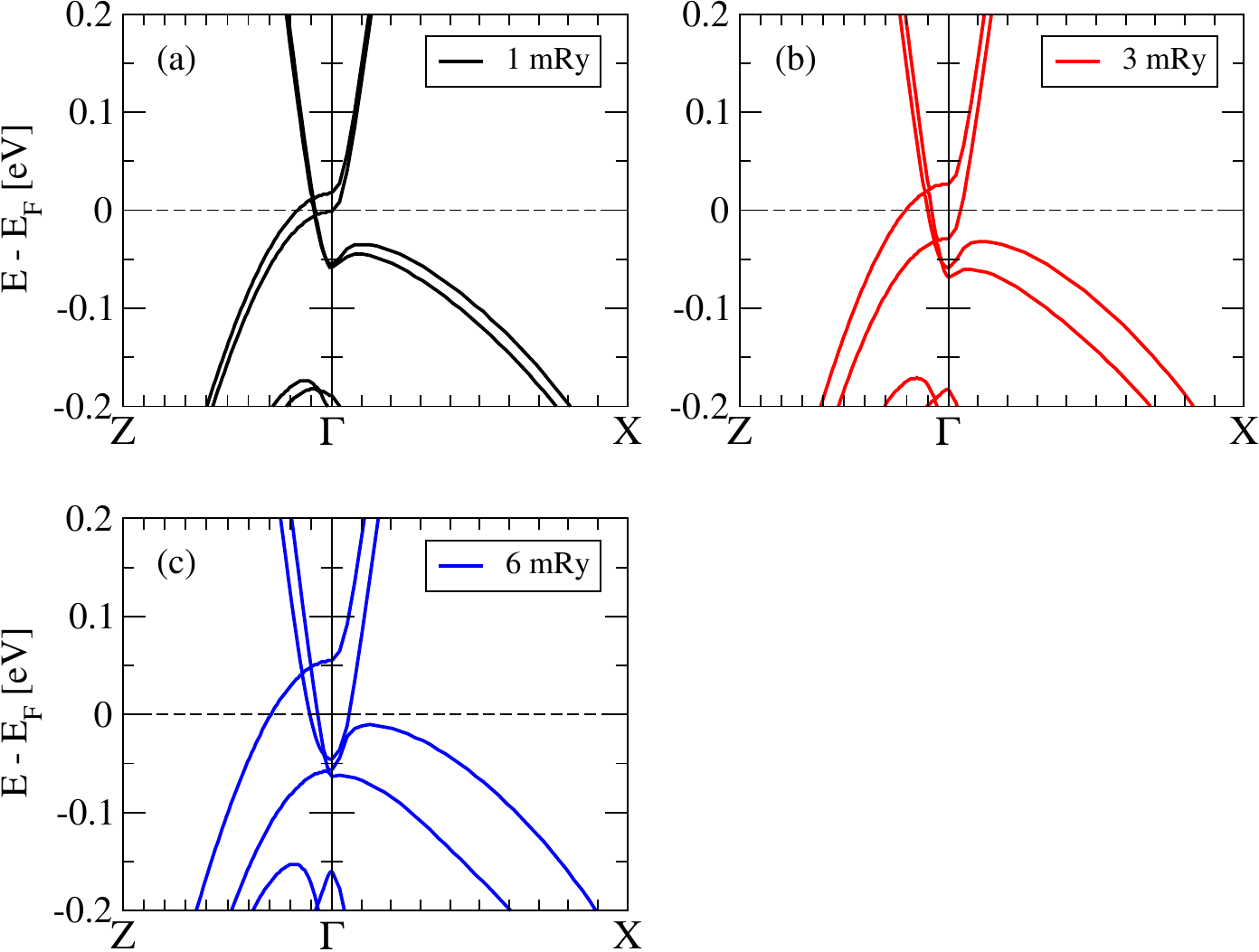}
\caption{\label{fig:cdasbands_Bext}
QS$GW$ electronic band structure of pristine Cd$_3$As$_2$ in the presence of an external applied
magnetic field $B_\text{ext}$, oriented in the $z$-direction.
Energies, in eV, are measured from the Fermi level $E_F=0$ reference energy.
(a) $B_\text{ext}=$ 1 mRy (13.6 meV or 235.8 Tesla),
(b) $B_\text{ext}=$ 3 mRy (40.8 meV or 705.8 Tesla),
(c) $B_\text{ext}=$ 6 mRy (72.2 meV or 1249.1 Tesla) which bears strong resemblance with
the bands of Mn-doped Cd$_3$As$_2$.
}
\end{figure}

Figure \ref{fig:cdasbands_Bext} shows the QS$GW$ band structure of pristine Cd$_3$As$_2$ obtained
in the presence of an external applied magnetic field $B_\text{ext}$ oriented in the $z$-direction.
Upon application of the magnetic fields, bands are split due to the Zeeman effect, as expected.
A large external field, as large as 6 mRy (72.2 meV or 1249.1 Tesla), generated a Zeeman band splitting 
similar to what is obtained for the Mn-doped Cd$_3$As$_2$ system.
For that field value, the band structure of pristine Cd$_3$As$_2$ bears some strong resemblance 
with the bands of the Mn-doped crystal;
although missing the apparent gapped band around $E-E_F=25$ meV seen in the Mn-doped case  
(comparing Figures \ref{fig:qsgwbands}(d) and \ref{fig:cdasbands_Bext}(c) ).

\begin{table}
\begin{equation}\nonumber
\begin{split}
\begin{array}{|c|c|c|c|c|}
\hline
\hline
\text{Energy range} & k-\text{path}  & \text{QS}GW  & \text{LDA}  & \text{GGA}  \\
\hline
200 \text{ to } 600 \text{ meV} & Z\Gamma   & 9.2 - 10.6 & \sim 4.0 &   \\
                                & \Gamma X  & 7.5 - 9.1  & \sim 2.0 &   \\
\hline
10 \text{ to } 25 \text{ meV} & Z\Gamma \text{ down} & \sim 9.0 &   & \sim 5.9 \\
                              & Z\Gamma \text{ up}   & \sim 1.5 &   & \sim 1.5 \\
\hline
\hline
\text{Dirac point } k_D [\AA^{-1}]       &      & 0.029 &                     & 0.042 \\
\hline
\end{array}
\end{split}
\end{equation}
\caption{\label{table:vF}
Band velocities (in units of $10^5$ [m/s]) from QS$GW$ and DFT band structure calculations of 
pristine Cd$_3$As$_2$, for different $k$-paths and different energy ranges. And
value of the Dirac point position $(0,0,k_D)$ along $\Gamma Z$.
}
\end{table}

\subsection{Model Hamiltonian: electronic structure and external magnetic field effects}
\label{sec:estrucH8}

\begin{figure}
\centering
\includegraphics[width=40mm]{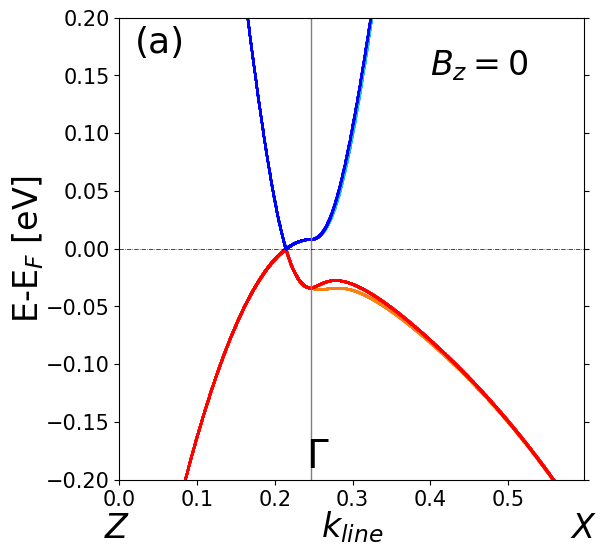}\includegraphics[width=42mm]{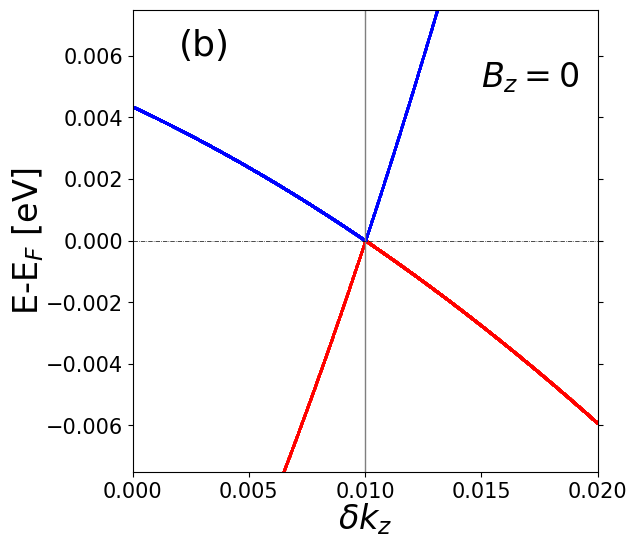}
\includegraphics[width=42mm]{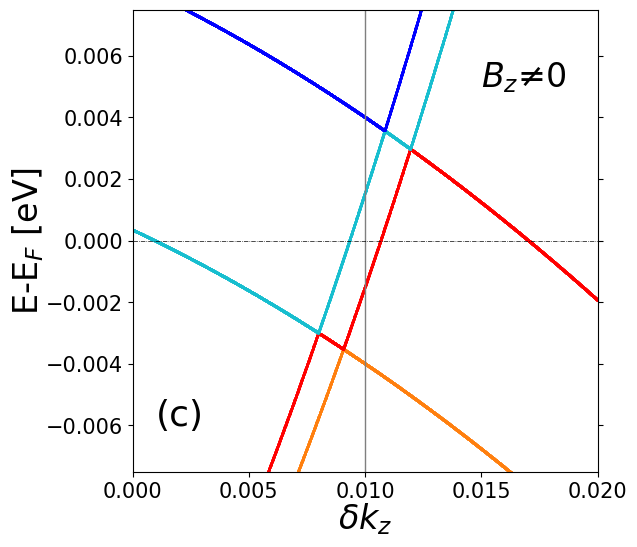}\includegraphics[width=40mm]{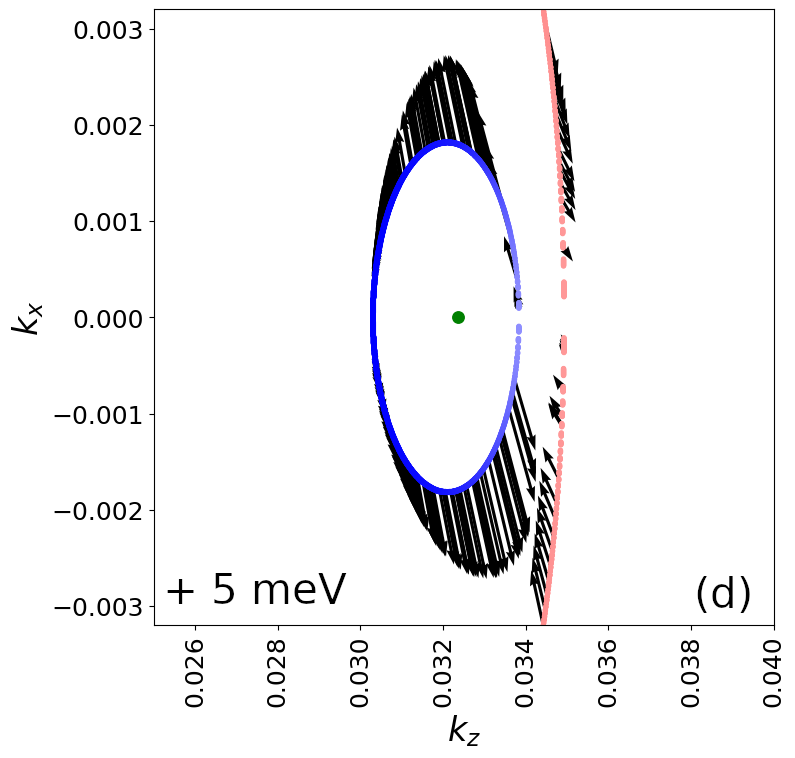}
\caption{\label{fig:kdotp_bands_Bz}
Band structure from Hamiltonian $H_8(\mathbf{k})$.
(a) No applied field. Bands are colored according the eigenvalues sorted
in ascending order. The $k$-path $k_\text{line}$ (in $\AA^{-1}$) goes from $Z$ (left) to 
$\Gamma$ (vertical black line), to $X$ (right) as in Fig.\ref{fig:qsgwbands}.
The Dirac point is along the $\Gamma Z$ line, located at $k_D \sim 0.032\ \AA^{-1}$
away from $\Gamma$.
(b) 
Zoom around the Dirac point along $k_z$ (in $\AA^{-1}$) with $k_z = k_D-0.01 +\delta k_z$.
(c) Splitting of the Dirac point with an applied field $\gamma B_z$ = 4 meV 
in the $z$-direction.
The original Dirac point is split into a
pair of Weyl points displaced away from $(0,0,k_D)$ and slightly shifted in energy around $E_F$.
(d) 
Two-dimensional constant-energy band structure in
the $k_z-k_x$ plane around the Dirac point (green dot), for a constant energy $E - E_F = 5$ meV.
The band coloring corresponds to the value of the spin component $S_z$, i.e. blue $S_z=-1$, red $S_z=+1$,
white $S_z=0$. The black arrows are the spin vectors $(S_y,S_x)$ in the $k_z-k_x$ plane.
}
\end{figure}

From \emph{ab-initio} calculations, we have seen that an applied field to the undoped Cd$_3$As$_2$ 
can generate similar bands to the Mn-doped Cd$_3$As$_2$ system. The next question to solve
is what happens to the Dirac points in the presence of Mn-doping and/or of the applied magnetic field?

To answer this question, we turn to a simpler model to describe the band structure of Cd$_3$As$_2$.
Using such a model allows us to explore the $k$-space bands more easily and to consider any arbitrary
magnitude and orientation of the magnetic field.
We consider a $k \cdot p$ model consisting of four orbitals ($s,p_{x,y,z}$) and two 
spins for each orbital,
and including a spin-orbit coupling term $H_\text{SO}$ as described in Ref~\cite{Wang:2013}.
Furthermore, we add a Zeeman-like term $H_\text{B}=-\gamma \mathbf{B} \boldsymbol{\sigma}$ where 
the magnetic field $\mathbf{B}=(B_x,B_y,B_z)$ couples to the spin Pauli 
matrices $\boldsymbol{\sigma}=(\sigma_x,\sigma_y,\sigma_z)$.
The Hamiltonian $H_8(\mathbf{k})$ (of size $8\times 8$) is a function of three components $k_x,k_y,k_z$
of the crystal momentum $\mathbf{k}$, and is fully described in Appendix \ref{app:kdotp}.
After diagonalization for each $\mathbf{k}$, the eigenvalues are used to build the corresponding band
structure and additional physical quantities as detailed below.
Such a model provides us with a quicker way to analyze how the Dirac points are displaced 
and/or split in $k$-space due to the application of the magnetic field.

Figure \ref{fig:kdotp_bands_Bz} shows the band structure obtained from the $k \cdot p$ $H_8$ Hamiltonian.
In the absence of a magnetic field, the model provides a fairly good representation of the QS$GW$
bands as can be seen by comparing Fig.~\ref{fig:kdotp_bands_Bz}(a) with the QS$GW$ bands Fig.~\ref{fig:qsgwbands}(c). 
{With the $k \cdot p$ model, we obtain a Dirac point at $k_D=0.032\ \AA^{-1}$,
very close to the  QS$GW$ calculations.
}

{
Upon application of a magnetic field in the $z$-direction, the Dirac point is split into individual Weyl points \cite{Baidya:2020} 
displaced away from $(0,0,k_D)$ and shifted symmetrically up and down in energy around $E_F$
as shown in Figure \ref{fig:kdotp_bands_Bz}(c). 
{
There are actually four band crossings, two above $E_F$ and two below.  
Two corresponding to a pair of ``conventional'' Weyl points (Chern number $\pm 1$) labelled $W^\pm$ 
in Fig.~\ref{fig:app_WandCpts_bz}, and two labelled $C^\pm$.  
Ref~\cite{Baidya:2020} concluded that the $C^\pm$ crossing points are "Weyl" points with Chern number 2.  
Our calculations align with Ref~\cite{Baidya:2020} regarding the $W^\pm$ points, but we do not find
pronounced Berry curvature around the $C^\pm$ points.  This is discussed further in Appendix \ref{app:D2W_bz}.
}

The identification of the Weyl bands and crossings can be made clearly by considering the bands spin texture.
The spin components $S_i$ ($i=x,y,z$) are calculated from the expectation value of the spin Pauli
matrices $S_i= \langle \sigma_i \rangle$ using the eigenstates of $H_8(\mathbf{k})$ at each
$\mathbf{k}$ point.
In Appendix \ref{app:D2W_bz}, we provide a full analysis of the spin texture of the different bands
and for different constant energies ranging from $E-E_F= 5$ meV to $E-E_F= -5$ meV. Our analysis
clearly identify the position of the Weyl points with opposite chirality.
Figure \ref{fig:kdotp_bands_Bz}(d) shows the two-dimensional constant-energy band structure,
in the $k_z-k_x$ plane for a constant energy $E-E_F = 5$ meV. 
The blue ellipse corresponds to a slice in the Weyl ``cone'' at $E-E_F > 0$, located around
the original Dirac point (green dot in Fig.~\ref{fig:kdotp_bands_Bz}(d)).
As discussed in the previous section, the bands around the Dirac points are anisotropic, i.e. they
disperse with different velocities for different $k$ orientation.
Hence a constant energy slice in the  Weyl ``cone'' has the shape of an ellipse, instead of a
circle (were the velocity of the Weyl cone to be isotropic).

For $E-E_F = 5$ meV, the constant energy ellipse has the spin component $S_z$ oriented
in one direction (opposite to the applied magnetic field) and anisotropic spin vectors $(S_y,S_x)$ 
oriented outwards the ellipse in the $k_z-k_x$ plane.
The outer (red) band ($k_z > 0.034\ \AA^{-1}$) shows a spin component $S_z$ with the opposite 
direction and a different kind of spin texture.
As detailed in Appendix \ref{app:D2W_bz}, the outer band corresponds to the second Weyl ``cone''
located below $E_F$ around the original Dirac point.
} 

Realistic band structure for real materials contains richer information in contrast to 
simpler band models considering only linear $k$-dispersion for the Weyl cones. The former
may lead to transport properties which differ from those calculated from simpler bands
(considering the change of velocity, spin texture with respect to change
of Fermi enery or temperature).

\begin{figure}
\centering
\includegraphics[width=42mm]{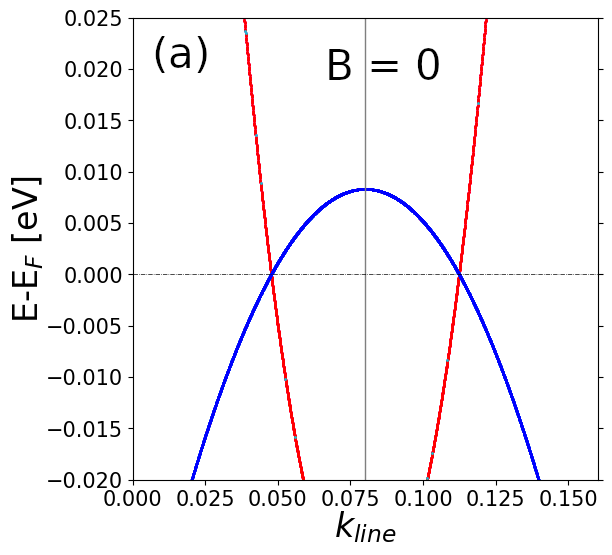}\includegraphics[width=42mm]{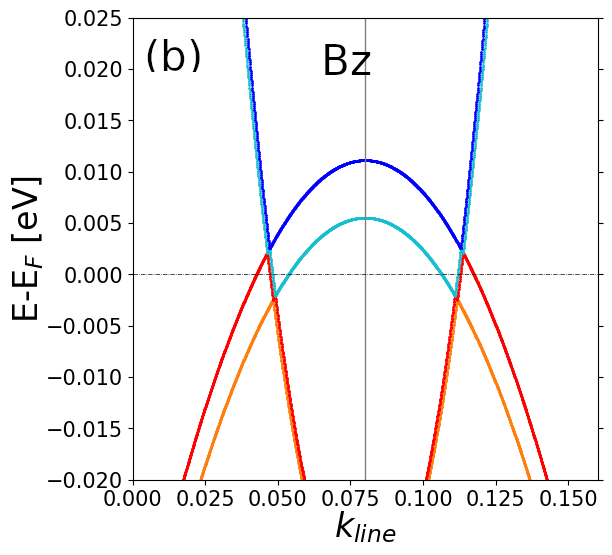}
\includegraphics[width=42mm]{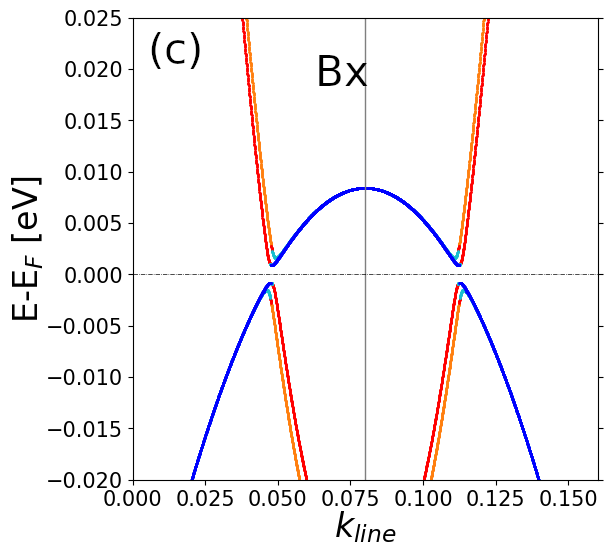}\includegraphics[width=42mm]{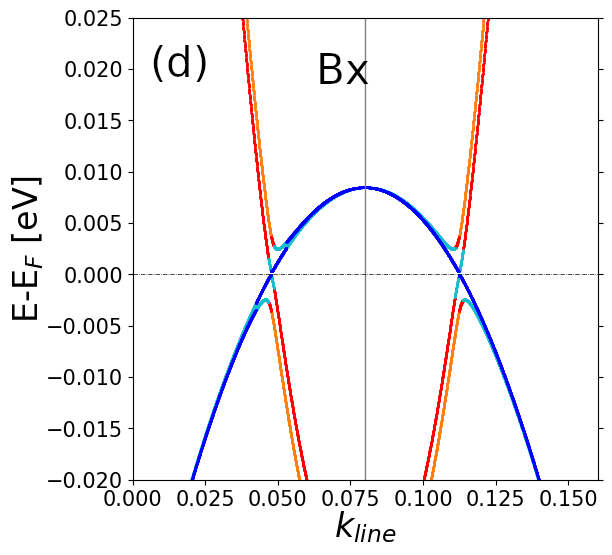}
\caption{\label{fig:kdotp_bandsBx}
Band structure from Hamiltonian $H_8(\mathbf{k})$.
The $k$-path $k_\text{line}$ is along the [001] direction and goes from
$(0,0,-0.08)$ to $(0,0,+0.08)\ \AA^{-1}$. 
(a) No applied field.
(b) Applied field in the $z$-direction $\gamma B_z$ = 2.8 meV.
(c) Applied field in the $x$-direction $\gamma B_x$ = 2.8 meV, $B_z=B_y=0$.
There is an apparent gap along the $k$-path at the Dirac points.
(d) Applied $B_x$ field and shifted $k$-path going from
$(+\delta k_x,0,-0.08)$ to $(+\delta k_x,0,+0.08) \AA^{-1}$.
The original Dirac points at $(0,0,\pm k_D)$ split into two pair
of Weyl points (crossing of 2 bands) at $(\pm \delta k_x,0,\pm k_D)$
with $\delta k_x = 0.0015\ \AA^{-1}$.
}
\end{figure}

So far we have considered the case of a magnetic field applied in the $z$-direction.
{
It is crucial to know how the Dirac points are affected by a changing orientation of the field.  
Figure \ref{fig:kdotp_bandsBx} compares how the Weyl points evolve from the Dirac points for fields 
along the $x-$ and $z-$direction.  
For $\mathbf{B} \parallel z$, bands drawn along the $(0,0,k_z)$ line, show that the Weyl points split in both $k$ 
and energy, as shown in Figure \ref{fig:kdotp_bands_Bz}.  The separation between Weyl points is parallel 
to $\mathbf{B}$.  
When $\mathbf{B} \parallel x$, a gap appears on this same line; however, the Dirac point splits into two Weyl points 
at $(\pm\delta{k}_x,0,\pm k_D)$. Fig. \ref{fig:kdotp_bandsBx}(c) depicts bands for the same excursion in
${k}_z$ but offset by a fixed $\delta{k}_x$.  
In general the connecting vector between Weyl points is parallel to $\mathbf{B}$.  
However, the perturbation under a $B_x$ field causes no energy splitting of the Weyl points, unlike the 
$\mathbf{B} \parallel z$ case.
}

\begin{figure}
\centering
\includegraphics[width=42mm]{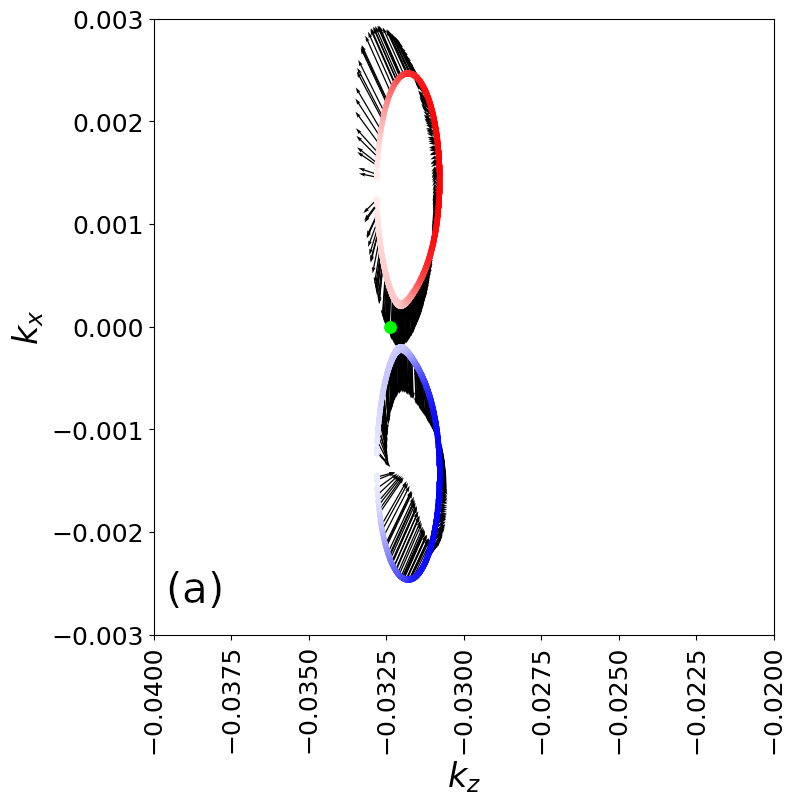}\includegraphics[width=42mm]{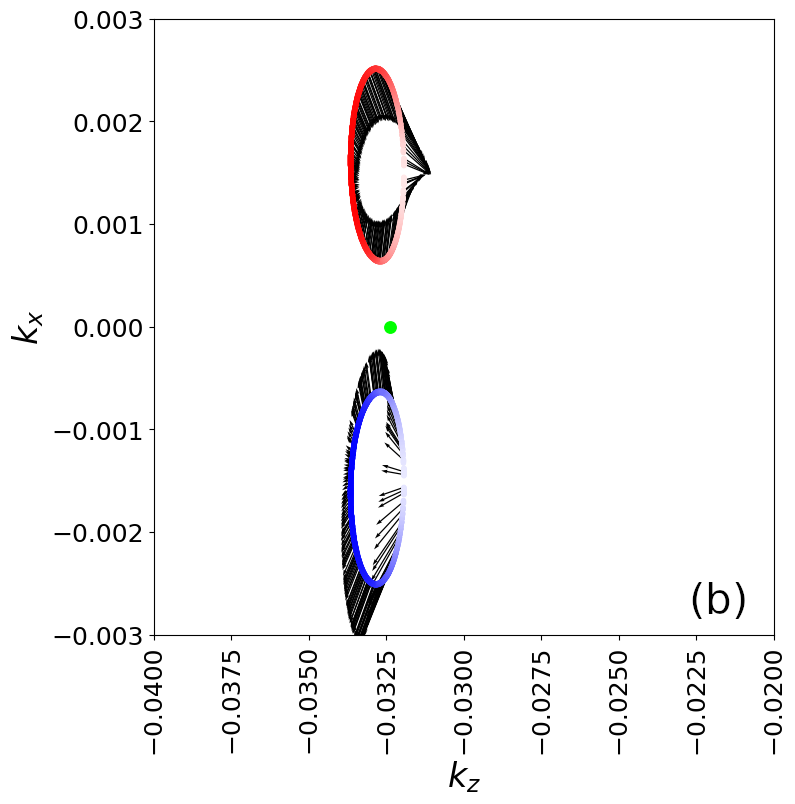}
\caption{\label{fig:kdotp_bandsBx_2Dspintext}
Two-dimensional constant-energy band structure from the $H_8(\mathbf{k})$ Hamiltonian in
the $k_z-k_x$ plane around a pair of Weyl points located at $(\pm \delta k_x,0,-k_D)$.
The applied field is $\gamma B_x$ = 2.8 meV, $B_z=B_y=0$.
The coloring code correspond to the spin component $S_z$, i.e. blue $S_z=-1$, red $S_z=+1$,
white $S_z=0$. The small arrows represents the spin vector $(S_y,S_x)$ in the $k_z-k_x$ plane.
The position of the green dot is $(0,0,-k_D)$.
(a) 
Constant energy $E$=+0.9 meV.
(b) 
Constant energy $E$=-0.6 meV.
The opposite spin texture (chirality) between the two Weyl points is clear.
Also note the change of sign of the spin texture $(S_y,S_x)$ for energy
above and below the Weyl crossing points.
}
\end{figure}

Figure \ref{fig:kdotp_bandsBx_2Dspintext} shows constant-energy band in the two-dimensional
$k_z-k_x$ plane around the pair of Weyl points located at $(\pm \delta k_x,0,-k_D)$,
moved along the $x$-direction of the field, in $k$-space.
The spin texture is also represented for the three spin components $S_{x,y,z}$.
The opposite chirality between the two Weyl points is clearly seen, as
well as the change of sign of the spin texture for energy
above and below the Weyl crossing points.

\subsection{Reconciling the doped and undoped Cd$_3$As$_2$ cases}
\label{sec:backtogw}

Our electronic structure calculations with the model Hamiltonian
have shown that the application of an external magnetic field splits the Dirac point into 
a pair at Weyl points of opposite chirality. 
They are displaced symmetrically around $kD$, possibly with a small energy shift.

\begin{figure}
\centering
\includegraphics[width=40mm,height=35mm]{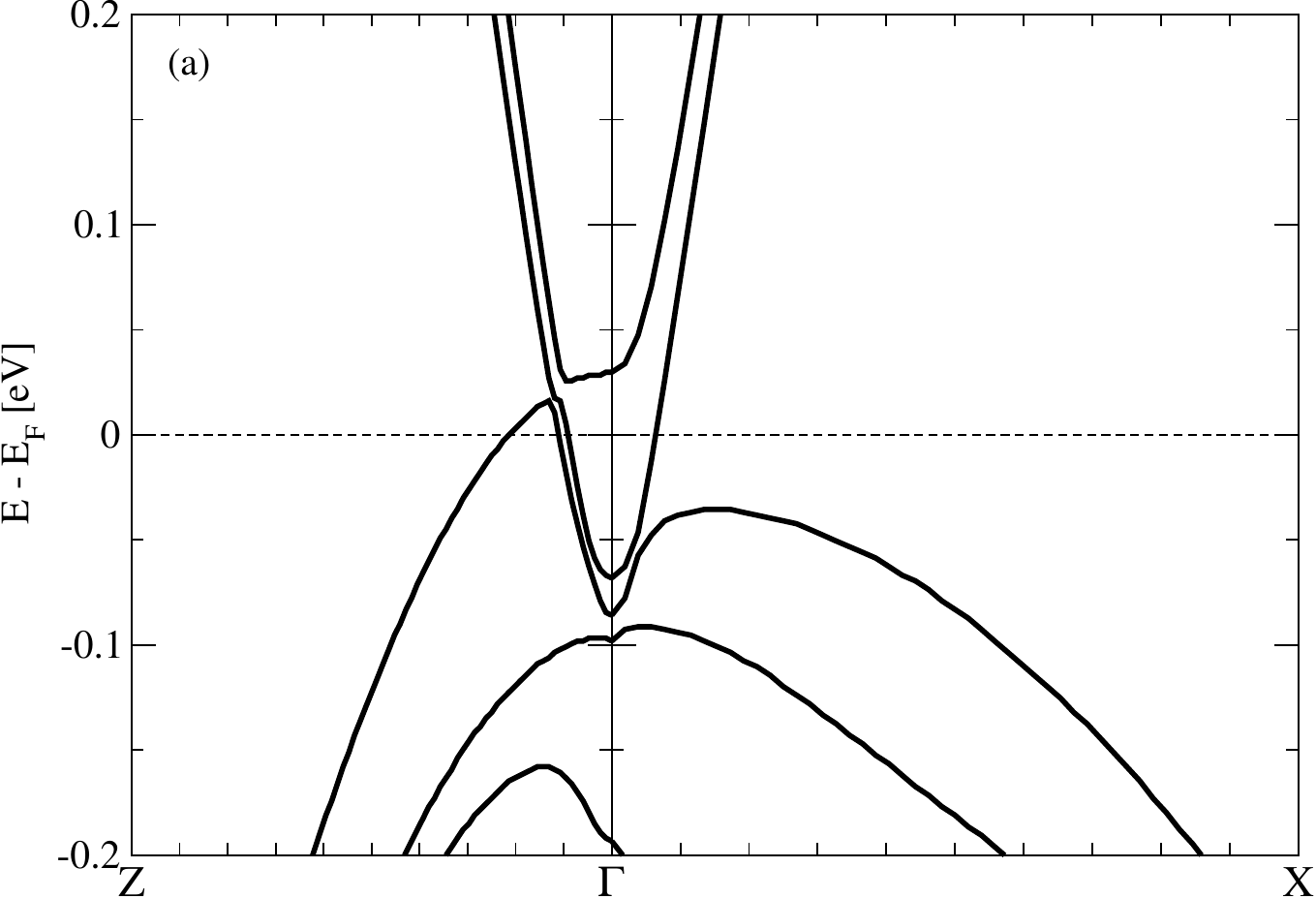}
\includegraphics[width=40mm,height=35mm]{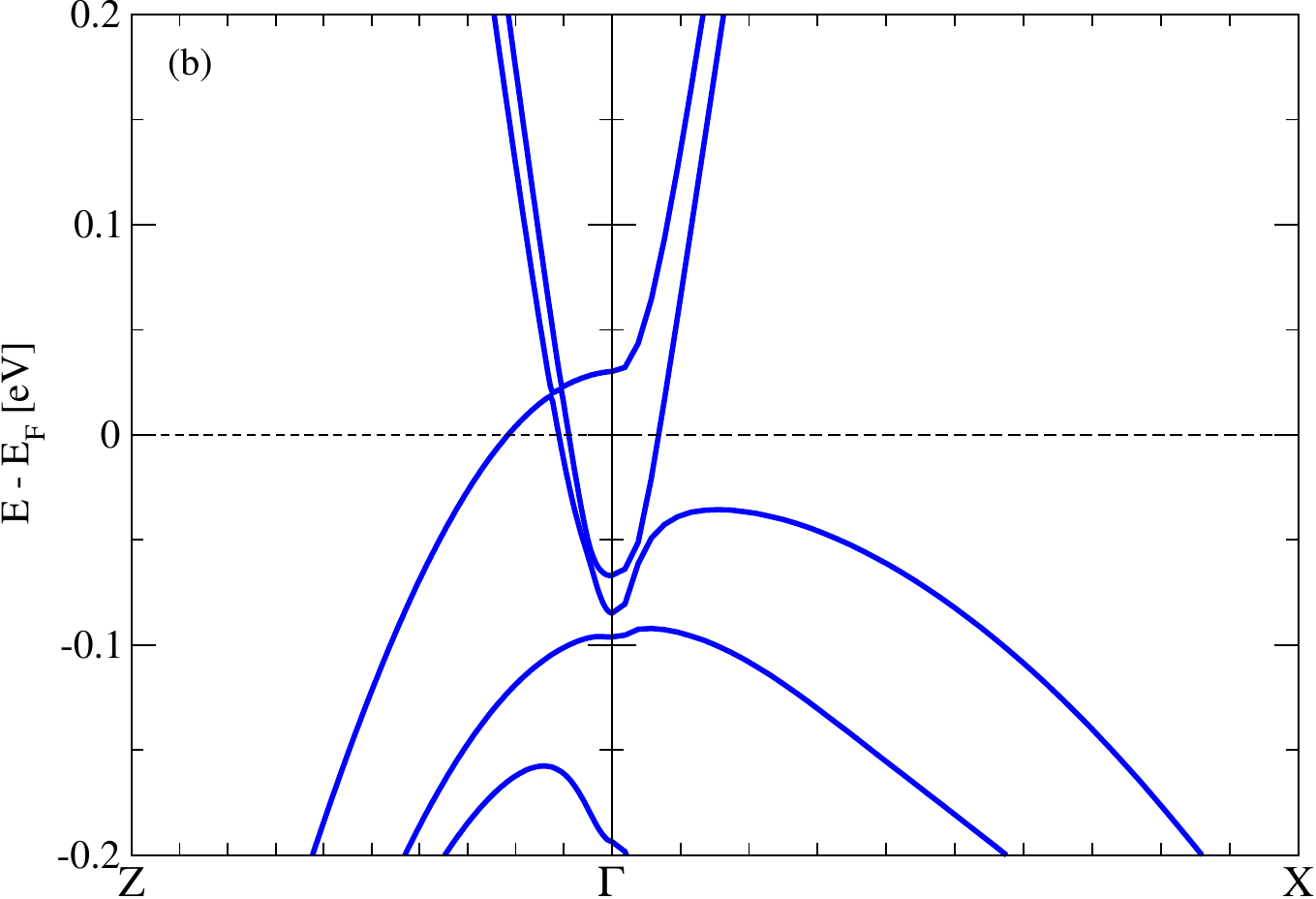}
\caption{\label{fig:qsgwbands_dkx}
Panel (a):
QS$GW$ electronic band structure of Cd$_3$As$_2$ for high symmetry $k$-space path $Z \Gamma X$. Same
as panel (d) in Fig.~\ref{fig:qsgwbands}.
Panel (b):
Bands along the $Z \Gamma X$ $k$-space path slightly shifted in the $k_x$ direction by
$\delta k_x = -0.0027\ \AA^{-1}$ for which the gap is closed. 
Energies, in eV, are measured from the Fermi level $E_F=0$
The bands in panel (b) look extremely similar to the bands in panel (c) in  Fig.~\ref{fig:cdasbands_Bext}.
}
\end{figure}

Such a behaviour should also, in principle, hold for the band structure of the realistic
cases of doped and undoped Cd$_3$As$_2$.
{In Mn-doped Cd$_3$As$_2$, 
the gap around $E-E_F=25 \text{meV}$ close to $\Gamma$ is only apparent: the bands do still cross 
but at a displaced $k_x$ point}.
Indeed, by plotting the QS$GW$ band structure of Mn-doped Cd$_3$As$_2$ along the $Z\Gamma X$
$k$-path slightly shifted in the $k_x$ direction, one recovers a closed gap as shown in
Figure \ref{fig:qsgwbands_dkx}. This is where one of the Weyl point has been moved by the
field generated by the magnetic impurity.

This implies that the local magnetic field due to the Mn impurity is slightly canted
away from the $z$-direction. This small deviation of the field towards the $x$-direction is most
probably related to 
spin-orbit coupling in the QS$GW$ Hamiltonian of Mn-doped Cd$_3$As$_2$ 
{generating a slight canting away from the $z$-axis.
}

Thus, the magnetic impurity does not open a gap in the band structure
and one can reasonably well emulate the ``exact'' band structure of Mn-doped Cd$_3$As$_2$
from the band structure of pristine Cd$_3$As$_2$ by applying an appropriate external
magnetic field. This is clearly seen by comparing Figure \ref{fig:qsgwbands_dkx}(b)
with Figure \ref{fig:cdasbands_Bext}(c). 

These results also lead us to believe that one can use the $k \cdot p$ model Hamiltonian, complemented
with the Zeeman term for an applied field, to 
calculate physical properties measured on Mn-doped Cd$_3$As$_2$ sample.

\subsection{Conductivity and orientation of the magnetic field}
\label{sec:conduc}

The transport properties of TSMs are intimately related to their band structures.
We have seen how the band structure of Cd$_3$As$_2$ is modified by magnetic impurity
doping and/or by an applied magnetic field.
We now turn to studying some effects of a magnetic field on the conductivity of bulk Cd$_3$As$_2$.

In Weyl TSM, an anomalous Hall DC conductivity is observed with a value lying between zero (conductivity of insulators)
and one quantum of conductance for topological insulator \cite{Burkov:2011,Panfilov:2014,Armitage:2018}.
The DC Hall conductivity, governed by Berry phase, is proportional to the separation in $k$-space of two Weyl points
which itself may depend on the applied magnetic field.

{Magnetotransport of bulk TSM involve additional features not included here (especially the role 
Landau levels play) and will be discussed elsewhere.
In the following we focus on how an applied field affects the symmetry of the dynamical conductivity tensor.}

From the eigenstates of $H_8(\mathbf{k})$ and from the velocity 
matrices $\nabla_{k_\alpha}H_8(\mathbf{k})$, we can calculate the 
conductivity tensor $\sigma_{\alpha\beta}(\omega)$ 
($\alpha,\beta=x,y,z$) within linear response theory \cite{Bastin:1971,Crepieux:2001,Bruus:2004,Vasko:2005,Morimoto:2012,Gradhand:2012}.
Detail of the calculations are provided in Appendix \ref{app:sigma}.
In the DC regime, the definition of $\sigma_{\alpha\beta}$ from the Berry curvature \cite{Gradhand:2012} 
implies that conductivity tensor is anti-symmetric, i.e. $\sigma_{\alpha\beta}=-\sigma_{\beta\alpha}$ 
for $\alpha \ne \beta$.
This is due to the property of the cross product of the current operator matrix elements entering
the definition  of the Berry curvature \cite{Gradhand:2012}.

In the AC regime ($\omega\ne 0$), the anti-symmetry of the conductivity tensor is not obvious, 
as shown in Appendix \ref{app:sigma}. The symmetry of the conductivity tensor will depend
on the intrinsic symmetry of the eigenstates, and therefore on the orientation of the magnetic field.
In the following, we show the symmetry of the off-diagonal components 
$\sigma_{\alpha\beta}(\omega)$ ($\alpha \ne \beta$)
for arbitrary orientation of the external magnetic field.

As a first step, we have checked that our calculations satisfy the Onsager relationships, as
expected, i.e.
$\sigma_{xy}(\omega;B_z) = \sigma_{yx}(\omega;-B_z)$
(and likewise for the $xz,zx$ components with an applied $B_y$ field, and for  
the $yz,zy$ components with an applied $B_x$).
Different symmetry relationships may hold for arbitrary orientations of the applied field.

The results of our calculations with different field orientations 
are summarised in Figure \ref{fig:symsigmab}.
For an applied field along the high symmetry Cartesian directions $x,y,z$ , only the (Hall) conductivity 
in the plane perpendicular to the field is non-zero, as should be expected. 
For example, with  $\mathbf{B}=(0,0,B_z)$, only the components  $\sigma_{xy,yx}$ are non-zero;
and similarly for the components $xz,zx$ ($yz,zy$) components with an applied $B_y$ ($B_x$).
These pairs of components are anti-symmetric, i.e. $\sigma_{\alpha\beta}(\omega)=-\sigma_{\beta\alpha}(\omega)$.
similarly to the DC case.

For other orientations of the field, more than one pair of ($\alpha\beta,\beta\alpha)$ components
have non-zero conductivity values. 
{This comes essentially from the symmetry of the rotation of the frame-axis and
the fact that only the ``perpendicular to the field'' (non-diagonal) components of the conductivity are non-zero.
}
For field orientation along some symmetry lines [110], [101], [011], 
we find that there is always at least one pair of $\alpha\beta$ components which provide anti-symmetry.
This is a very interesting outcome for the optical properties of TSM \cite{Kotov:2016,Hofmann:2016,Kotov:2018}.
Indeed,
the anti-symmetry of the (dynamical) conductivity tensor is a central key in order to obtain non-reciprocity effects of 
light \cite{Kotov:2018} 
and thermal radiation  \cite{Zhao:2020} in Weyl TSMs.

{Our results imply that the local field, whatever its origin, does not need to be
strictly oriented along high symmetry Cartesian axis to realize antisymmetry of the optical conductivity tensor, 
and therefore in order to obtain non-reciprocal effects in Weyl TSMs.
}.

\newcommand{\vvec}[3]{\left(\begin{array}{c} #1 \\ #2 \\ #3 \end{array}\right)}
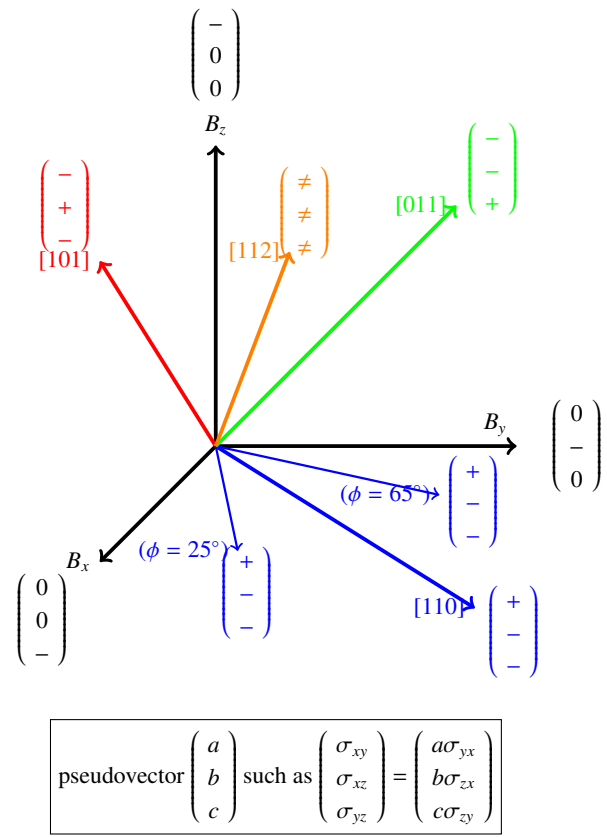
\begin{figure}
\begin{center}
\begin{tikzpicture}[scale = 0.8]
 \draw[line width=.5 mm, ->,black] (0,0,0) -- (5,0,0) node[above left] {$B_y$}  node[pos=1.2]{$\vvec{0}{-}{0}$} ; 
 \draw[line width=.5 mm,->,black] (0,0,0) -- (0,5,0) node[above] {$B_z$}	node[pos=1.3]{$\vvec{-}{0}{0}$} ; 
 \draw[line width=.5 mm,->,black] (0,0,0) -- (0,0,5) node[left] {$B_x$}   node[pos=1.5]{$\vvec{0}{0}{-}$} ; 
 \draw[line width=.5 mm,->,red] (0,0,0) -- (0,5,5) node[left] {$[101]$} node[pos=1.3]{$\vvec{-}{+}{-}$} ;
 \draw[line width=.5 mm,->,blue] (0,0,0) -- (7,0,7) node[left] {$[110]$} node[pos=1.15]{$\vvec{+}{-}{-}$}; 
 \draw[line width=.3 mm,->,blue] (0,0,0) -- (4.53,0,2.11) node[left] {$(\phi=65^\circ)$} node[pos=1.15]{$\vvec{+}{-}{-}$}; 
 \draw[line width=.3 mm,->,blue] (0,0,0) -- (2.11,0,4.53) node[left] {$(\phi=25^\circ)$} node[pos=1.4]{$\vvec{+}{-}{-}$}; 
 \draw[line width=.5 mm,->,green] (0,0,0) -- (4,4,0) node[left] {$[011]$} node[pos=1.15]{$\vvec{-}{-}{+}$} ;
 \draw[line width=.5 mm,->,orange] (0,0,0) -- (2,4,2) node[left] {$[112]$} node[pos=1.2]{$\vvec{\ne}{\ne}{\ne}$} ;
 \node[draw] at ( 1,-5.5) {\text{pseudovector} $\vvec{a}{b}{c}$ \text{such as} $\vvec{\sigma_{xy}}{\sigma_{xz}}{\sigma_{yz}} = 
 \vvec{a \sigma_{yx}}{b \sigma_{zx}}{c \sigma_{zy}}$ };  
%
\end{tikzpicture}
\end{center}
\caption{\label{fig:symsigmab}
Symmetry of the off-diagonal components of the AC conductivity tensor
$\sigma_{\alpha\beta}(\omega)$. For a given field orientation, the symmetry relationships 
hold for all energy $\omega \ne 0$.
The different arrows represent the direction $[hkl]$ of the applied magnetic field
in the 3D real space.
The pseudovector gives the relationship between 
the 3 pairs $(xy,yx)$, $(xz,zx)$ and $(yz,zy)$ of the conductivity components. 
}
\end{figure}

\section{Conclusion}
\label{sec:ccl}

We have studied electronic structure of bulk Cd$_3$As$_2$
by using QS$GW$ electronic structure calculations  
and a $k \cdot p$ model Hamiltonian.
We have shown that substitutional doping by Mn impurities, in the small doping regime,
induce a topological phase transition, i.e. the original Dirac points in pristine Cd$_3$As$_2$ 
split into individual Weyl points.
We have demonstrated that the electronic structure of Mn-doped Cd$_3$As$_2$ can be fairly well
reproduced by the electronic structure of pristine Cd$_3$As$_2$ with an appropriate external 
magnetic field. 
This result highlights opportunities for unique device functionality based on band structure tuning 
which is not found in conventional magnetic Weyl TSM.
Using a $k \cdot p$ model Hamiltonian, in the presence of an applied magnetic field, we have studied
the conductivity of bulk Cd$_3$As$_2$ for different orientations of the applied field and shown
the symmetry properties of the conductivity tensor.

{From an experimental standpoint, substitutional doping with Mn must first be demonstrated in Cd$_3$As$_2$
\cite{ARice:2025}, followed by verification of changes in the bandstructure and properties. 
}

{At dilute doping, spins will be disordered; causing the net field to be small.  However an applied field may act to
align the Mn spins, greatly enhancing the spin susceptibility relative to pristine Cd$_3$As$_2$.
At higher doping the Mn can interact, and affect the spin ordering and modify the effective field.  
Also we considered substitutional doping only.  The interstitial Mn(I)
may act differently from the substitutional Mn(s), especially as Mn(I) will dope the system.  Also Mn(I)-Mn(s)
interactions can be very different from Mn(s)-Mn(s) interaction, which can affect the thermodynamic distribution of
dopants and the total spin.
}

\begin{acknowledgements}

This work was authored in part by the National Renewable Energy
Laboratory for the U.S. Department of Energy (DOE) under Contract
No. DE-AC36-08GO28308. Funding was provided by the U.S. Department of
Energy (DOE), Office of Science, Basic Energy Sciences, Physical
Behavior of Materials Program as part of the “Disorder in Topological
Semimetals” project. The views expressed in the article do not
necessarily represent the views of the DOE or the U.S. Government. The
U.S. Government retains and the publisher, by accepting the article for
publication, acknowledges that the U.S. Government retains a
nonexclusive, paid-up, irrevocable, worldwide license to publish or
reproduce the published form of this work, or allow others to do so, for
U.S. Government purposes.  We acknowledge the use of the National Energy
Research Scientific Computing Center, under Contract
No. DE-AC02-05CH11231 using NERSC award BES-ERCAP0021783 and we also
acknowledge that a portion of the research was performed using computational 
resources sponsored by the Department of Energy’s Office of
Energy Efficiency and Renewable Energy and located at the National
Renewable Energy Laboratory. HN acknowledges financial support from
NREL via a subcontract agreement between KCL and NREL.

\end{acknowledgements}

\appendix

\section{Questaal calculations}
\label{app:questaal}

For electronic structure calculations, we use the Quasiparticle Self-Consistent $GW$ 
approximation \cite{Faleev:2004,MvS:2006,Kotani:2007}.
Quasiparticlization of the $GW$ self-energy yields a static one-particle band similar in form to Hartree–Fock or DFT, 
but its fidelity is high because the potential is constructed so that energy bands physically correspond to excitation 
energies. 
This is because the QS$GW$ self-consistency condition causes the poles of the noninteracting Green’s function to 
coincide with the interacting one. 
DFT energy bands, by contrast, have no physical meaning even though they are widely interpreted as excitation
energies.
{See Refs.~\cite{Pashov:2020,questaal:web} for further details on the Questaal code and implementation of QS$GW$.
}

We use 
a primitive unit cell, consisting of 80 atoms 
(48 atoms of Cd and 32 of As), with broken inversion symmetry.
The lattice cell vectors are $(-a,a,b) ; (a,-a,b) (a,a,-b)$ for the pristine Cd$_3$As$_2$ system
where
$a = 6.313747\ \AA$ and $b = 12.766886\ \AA$.
For the Mn-doped case, one Cd atom is exchanged with one atom of Mn, i.e. corresponding to a 
doping of 1/48 $\sim$ 2 \%
(an equivalent supercell built from the vectors $(2a,0,0); (0,2a,0); (a;a;b)$ is used).

For the DFT calculations, a $k$-point grid of $(4 \times 4 \times 4)$ was used for the DFT calculations.
A smaller grid of $(3 \times 3 \times 3)$ was used for QS$GW$ calculations. 
The self-consistency of the DFT/QS$GW$ calculations is achieved when the 
RMS change in output-input self-energy and charge density were both converged to 
about $6 \times 10^{-6}$.
For QS$GW$ calculations, all the states are used in the present
case (5664 states in the total system, with $\sim$ 4612 of them
unoccupied). Questaal’s basis set is tailored to the potential \cite{Pashov:2020}, so that quasiparticle 
levels converge much more rapidly with the number of states than what occurs for 
a plane wave basis.
Band structure calculations, in the presence of the $GW$ self-energy, were performed 
with a denser k-point grid mesh.

\section{The $k \cdot p$ Hamiltonian}
\label{app:kdotp}

The $k \cdot p$ model Hamiltonian \cite{Wang:2013,Smith:2024} is build on 
4 bands ($s-$ and $p-$ orbitals) and 2 spins, including spin-orbit coupling (SOC) and a Zeeman-like term:
\begin{equation}
\begin{split}
H_8(\mathbf{k}) = H_4(\mathbf{k}) \otimes \mathbb{I}_\sigma + H_\text{SO} + H_\text{Z} \ .
\end{split}
\label{eq:H8x8}
\end{equation}
where $\mathbb{I}_\sigma$ is the identity in the spin space.

For a given spin, the Hamiltonian $H_4(\mathbf{k})$, in the basis of the $s-$ and $p-$ orbitals, is given by
\onecolumngrid 
\begin{equation}
\begin{split}
H_4(\mathbf{k}) =
\left[
\begin{array}{cccc}
A k^2 + E_s	& ik_x P_x	& ik_y P_y	& ik_z P_z + d_\text{IS} 		\\
-i k_x P_x	& L k_x^2 + M(k_y^2 + k_z^2) + E_{px}	& N k_x k_y	& N k_x k_z	\\
-i k_y P_y	& N k_x k_y	& L k_y^2 + M(k_x^2 + k_z^2) + E_{py}	& N k_y k_z	\\
-i k_z P_z +d_\text{IS}	& N k_x k_z	& N k_y k_z	& L k_z^2 + M(k_x^2 + k_y^2) + E_{pz}	\\
\end{array}
\right]
\end{split}
\label{eq:H4k}
\end{equation}
\twocolumngrid 

Cubic symmetry is broken by using: $E_{px}=E_{py}=E_p$ and $E_{pz} = E_p - \delta$.
The parameter $d_\text{IS}$ is introduced to break the inversion symmetry \cite{Wang:2013}.

The SOC term is
\begin{equation}
\begin{split}
H_\text{SO} =\frac{\Delta}{2} \mathbf{L}\cdot\boldsymbol{\sigma} =
\frac{\Delta}{2} \left[
\begin{array}{cc}
L_z	& L_- 		\\
L_+	& -L_z
\end{array}
\right]
\end{split}
\label{eq:HSO}
\end{equation}
where $\boldsymbol{\sigma}$ is the vector of Pauli matrices $\sigma_\alpha$,
$L_\alpha$ are (4$\times$4) matrices in the $sp-$ orbital basis,
and
$L_\pm = L_x \pm i L_y$with 

For the $s-$ orbital, the angular momentum is zero. Hence we have
\begin{equation}
\begin{split}
L_\alpha = 
\left[
\begin{array}{cc}
0 & 0_{3\times 1} \\
0_{1\times 3} & l_\alpha \\
\end{array}
\right] 
\end{split}
\label{eq:Lxyz}
\end{equation}
In the basis of the $p-$ orbitals, the angular momentum is given by
\begin{equation}
\begin{split}
l_x = i\hbar
\left[
\begin{array}{ccc}
0 & 0 & 0 \\
0 & 0 & -1 \\
0 & 1 & 0 \\
\end{array}
\right] , \ 
l_y = i\hbar
\left[
\begin{array}{ccc}
0 & 0 & 1 \\
0 & 0 & 0 \\
-1 & 0 & 0 \\
\end{array}
\right] , \ 
l_z = i\hbar
\left[
\begin{array}{ccc}
0 & -1 & 0 \\
1 & 0 & 0 \\
0 & 0 & 0 \\
\end{array}
\right]
\end{split}
\label{eq:lxyz}
\end{equation}

The Zeeman term is
\begin{equation}
\begin{split}
H_\text{Z} = -\gamma \mathbf{B}\cdot\boldsymbol{\sigma} 
= -\gamma
\left[
\begin{array}{cc}
B_z	& B_- 		\\
B_+	& -B_z
\end{array}
\right]
\end{split}
\label{eq:HBext}
\end{equation}
with
$B_\pm = B_x \pm i B_y$.

In the $sp-$ orbital basis, we take
\begin{equation}
\begin{split}
B_\alpha = 
\left[
\begin{array}{cccc}
B_\alpha^s & 0 & 0 & 0    \\
0 & B_\alpha^{px} & 0 & 0 \\
0 & 0 & B_\alpha^{py} & 0 \\
0 & 0 & 0 & B_\alpha^{pz} \\
\end{array}
\right]
\end{split}
\label{eq:lxyz}
\end{equation}
and
apply the $B$ field ``uniformly'' on the $p-$ orbitals only, i.e.
$B_\alpha^{px}=B_\alpha^{py}=B_\alpha^{pz}=B_\alpha$ and $B_\alpha^s=0$.

Note that calculations with $B_\alpha^s=B_\alpha$ do not change significantly
the electronic structure around the Dirac point; and not affect the results
of the symmetry of the conductivity tensor upon the direction of the applied
field.

The value of the parameters are taken from Ref.~[\onlinecite{Wang:2013}]
with a small modification of the SO coupling constant (here we take
$\Delta=$ 0.155 eV) to make the Dirac points align 
with the Fermi level $E_F$ within a range of $10^{-5}$ eV.

\section{Linear response conductivity}
\label{app:sigma}

The (dynamical) electrical conductivity is obtained from linear response
theory and given by \cite{Bastin:1971,Bruus:2004,Vasko:2005}

\begin{equation}
\begin{split}
\sigma_{\alpha\beta}(\omega) = \sigma_{\alpha\beta}^g(\omega) + \frac{e^2}{i\omega} S_{\alpha\beta}(\omega)
\end{split}
\label{eq:sigmagen}
\end{equation}
where $\sigma_{\alpha\beta}^g(\omega)$ is the so-called gauge term proportional to $\delta_{\alpha\beta}$
\cite{Bastin:1971,Bruus:2004,Vasko:2005} and $\alpha,\beta=x,y,z$ the Cartesian coordinates.
The current-current correlation function $S_{\alpha\beta}(\omega)$ is 
\begin{equation}
\begin{split}
S_{\alpha\beta}(\omega) = -2
\int \frac{{\rm d}\epsilon}{2\pi} f_\epsilon \
{\rm Tr}
\left[
j_\alpha {\rm Im}G(\epsilon) j_\beta G_+(\epsilon+\hbar\omega)  \right. \\
\left.
+\ j_\alpha G_-(\epsilon-\hbar\omega) j_\beta {\rm Im}G(\epsilon)   
\right]
\end{split}
\label{eq:sigmaSab1b}
\end{equation}
where $f_\epsilon$ is the Fermi-Dirac distribution function, $j_{\alpha,\beta}$ the current operators,
and $G_\pm$ are the Green's functions
$G_\pm = (\epsilon - H_8(\mathbf{k}) \pm i\eta)^{-1}$ (with $\eta\rightarrow 0^+$) and 
${\rm Im}G(\epsilon)=\frac{1}{2i} (G_+ - G_-) = - \pi \delta(\epsilon - H_8)$.

Using a compact notation for the eigenstates of $H_8(\mathbf{k})$, 
$E_n \equiv E_n(\mathbf{k})$ and $\vert n \rangle \equiv \vert u_n(\mathbf{k})\rangle$,
one recovers the usual expression for 
the dynamical conductivity $\sigma_{\alpha\beta}(\omega)$
(with $\alpha \ne \beta$) \cite{Bastin:1971,Crepieux:2001,Bruus:2004,Vasko:2005,Morimoto:2012,Gradhand:2012}

\begin{equation}
\begin{split}
\sigma_{\alpha\beta}(\omega)  & = 
{ -i e^2 \hbar} 
\sum_{n,m \ne n } \left( \frac{f_{E_m} - f_{E_n}}{\hbar\omega} \right)
\frac{\langle n\vert j_\alpha\vert m \rangle \langle m\vert j_\beta\vert n\rangle}{(E_m-E_n) +\hbar\omega+i\eta} \\
& =
\frac{ e^2 } {i \omega}
\sum_{n,m \ne n } (f_{E_n} - f_{E_m})
\frac{\langle n\vert j_\alpha\vert m \rangle \langle m\vert j_\beta\vert n\rangle}{(E_n-E_m) -\hbar\omega-i\eta} 
\end{split}
\label{eq:sigmaAC_1}
\end{equation}

The matrix elements $\langle n\vert j_\alpha\vert m \rangle$ of the $\alpha$-component 
of the current operator $\bm{j}$ are obtained, in the $k$-space, from the matrix elements 
$\langle u_n(\bm{k})\vert \nabla_{k_\alpha}H_8(\mathbf{k})\vert u_m(\bm{k})\rangle$.
Analytical expressions for $\nabla_{k_\alpha}H_8(\mathbf{k})$ are readily obtained from Eq.(\ref{eq:H4k}).

Also, note that by swapping the $\alpha,\beta$ indices, one obtains an expression for
$\sigma_{\beta\alpha}(\omega)$ 
which is clearly different from Eq.~\ref{eq:sigmaAC_1}:
\begin{equation}
\begin{split}
\sigma_{\beta\alpha}(\omega)  & = 
\frac{ e^2 } {i \omega}
\sum_{n,m \ne n } (f_{E_m} - f_{E_n})
\frac{\langle n\vert j_\alpha\vert m \rangle \langle m\vert j_\beta\vert n\rangle}{(E_m-E_n) -\hbar\omega-i\eta} \\
& \ne \sigma_{\alpha\beta}(\omega) \\
\text{and } & \ne - \sigma_{\alpha\beta}(\omega)
\end{split}
\label{eq:sigmaAC_2}
\end{equation}
Therefore, unless there is a very specific symmetry in the Hamiltonian, reflected in the spectrum $E_n$ and the
eigenstates $\vert n\rangle$, one cannot easily conclude if the conductivity tensor is symmetric, or anti-symmetric.
{Also note that, $H_8(\mathbf{k})$ and $\nabla_{k_\alpha}H_8(\mathbf{k})$ being Hermitian, is not
enough to explain the (anti)symmetry.}

Finally,
when performing actual calculations,  
the sums implicitly include the $k$-dependence of the eigenstates, i.e.
\begin{equation}
\begin{split}
\sum_{n, m \ne n} \dots \rightarrow \int \frac{{\rm d}\bm{k}}{(2\pi)^3} \sum_{n=1,m \ne n}^{\text{8}} \dots \\
\end{split}
\end{equation}
Numerical calculations were performed for a regular cubic $k$-point grid, centered on
the $\Gamma$ point, ranging from -0.08 to 0.08 $\AA^{-1}$ in the three Cartesian directions.
A mesh of ($355 \times 355 \times 355$) was used, i.e. more than 44 millions $k$-points in order to 
fully capture the contributions of the Weyl points \cite{Gradhand:2012}.


\section{Indentification of the Weyl points}
\label{app:D2W_bz}

{
In this appendix, we show in detail the spin texture of the bands around the
Dirac point at $(0,0,k_D$) in Figure \ref{fig:kdotp_bands_Bz}.
We calculate the spin components $S_{x,y,z}$
from expectation value of the Pauli matrices $S_i= \langle \sigma_i \rangle$ using
the eigenstates of $H_8(\mathbf{k})$ at each $\mathbf{k}$ point.

In Figure~\ref{fig:app_D2W_bz}, the 2D maps represent the spin vectors $(S_y,S_x)$ in the
$k_z-k_x$ plane as black arrows. The color coding of the 2D constant energy contours
corresponds to the $S_z$ spin component: blue for minority spin $S_z=-1$, red for
majority spin $S_z=1$, and white for $S_z \sim 0$.  

Due to the field applied in the $z$-direction, the Dirac point is split
into individual Weyl points \cite{Baidya:2020} located at $(0,0,k_z = k_D \mp \delta k_W^\pm)$.
The constant energy slices in the Weyl ``cones'' have elliptic shapes because 
of the anisotropy of the band velocity in $k$-space.
The Weyl points are shifted in energy from $E_F$ by the Zeeman effect,
down by $-E_W^+$ for the Weyl point $W^+$ with majority spin along the $z$-direction ,
and  shifted up by $+E_W^-$ for the Weyl point $W^-$ with minority $S_z$ spin
directed in the opposite direction of the applied field (see red and blue ellipses
in Fig.~\ref{fig:app_D2W_bz} respectively).

Let us first focus on the Weyl point $W^-$ above $E_F$. 
For energies above the Weyl point (i.e. $E - E_F > +E_W^-$), 
the spin vectors $(S_y,S_x)$ are directed outwards the blue ellipses,
while for energies below the Weyl point the spin vectors $(S_y,S_x)$ are directed inwards.

The other Weyl point $W^+$, located below $E_F$, has the opposite chirality as expected.
That is for energies $E - E_F > -E_W^+$ above the Weyl point, 
the spin vectors $(S_y,S_x)$ are directed inwards the red ellipses,
while for energies below the Weyl point the spin vectors $(S_y,S_x)$ are directed outwards.

Note that the constant energy slices of a given Weyl ``cone'', i.e. blue or red ellipses
for $W^\mp$
(with Chern number $\pm 1$ according to Ref~\cite{Baidya:2020}),
are accompanied with another band (red or blue respectively) with opposite spin $S_z$ direction
which originates from the other Weyl ``cone'' $W^\pm$ (at opposite energy).

{
The two other crossings, labelled  $C^\pm$ in Figure \ref{fig:app_WandCpts_bz}(a), have been identified  as 
other Weyl points associated with higher Chern number in Ref~\cite{Baidya:2020}. 
Our results reflect a different picture for the $C^\pm$ points, and the difference can be important because
of the central role spin texture plays in transport properties of TSM. 
For  example, the (anomalous) Hall conductivity is given in terms of the Berry curvature \cite{Gradhand:2012,Baidya:2020}.
}

From the eigenstates of $H_8(\mathbf{k})$, we can calculate the $(\gamma = x,y,z)$ components of the Berry curvature  
$\Omega_{n,\gamma}$ for a given eigenstate $n$ as follows:

\begin{equation}
\begin{split}
\Omega_{n,\gamma}
& = i\hbar^2 \sum_{m \ne n} 
\frac{\langle n\vert j_\alpha\vert m \rangle \langle m\vert j_\beta\vert n\rangle - \langle n\vert j_\beta\vert m \rangle \langle m\vert j_\alpha\vert n\rangle}{(E_n-E_m)^2} \\
\end{split}
\label{eq:Bcurv_n}
\end{equation}
with $\gamma \ne \alpha, \beta$ and $\alpha \ne \beta$.

{
Figure \ref{fig:app_WandCpts_bz} shows two 3D maps, in the $k$-space, of the Berry curvature vectors
$(\bar\Omega_x,\bar\Omega_y,\bar\Omega_z)$ as blue arrows.
We have intentionally displayed the arrows as short arrows in order to avoid strong overlap of the different 
vectors at different $k$-points which may lead to confusion.
The quantities $\bar\Omega_\gamma$ come from the summation of $\Omega_{n,\gamma}$ over an chosen energy
window $E_n \in [E_\text{min},E_\text{max}]$.
We have choosen energy windows around the pairs $(W,C)^\pm$ to check which of the crossing points provide
the more Berry curvature.
In Fig.~\ref{fig:app_WandCpts_bz}(b), the summation is performed for the energy window $[+3,+12.5]\ \text{meV}$ around
the pair of $W^-$ and $C^-$ crossing points; 
in Fig.~\ref{fig:app_WandCpts_bz}(c) for the energy window $[-12.5,-3]\ \text{meV}$ around
the pair of $W^+$ and $C^+$ crossings. 
The line of symbols, at $k_x=k_y=0$, represent the sequence of $C^+,W^+,D,W^-,C^-$ points along $k_z$, with the
Dirac point $D$ at $(0,0,k_D)$. 
For both integration windows, one can clearly see a large contribution of Berry curvature around the Weyl points 
$W^\pm$ (black dots) in comparison to a smaller Berry curvature around the crossing points $C^\pm$ (green diamonds).
This tends to indicate that, amongst the four band crossings $(W,C)^\pm$ , the Weyl points $W^\pm$ will be playing a more 
important role in the transport properties based on Berry curvatures.
}

{
We have not attempted to analyze why Ref \cite{Baidya:2020} and this work arrive at different conclusions 
about the $C^\pm$ points.  
The discrepancy can probably be traced to a difference in the construction of the k.p hamiltonians.
}

\onecolumngrid 

\begin{figure}
\centering
\includegraphics[width=37mm]{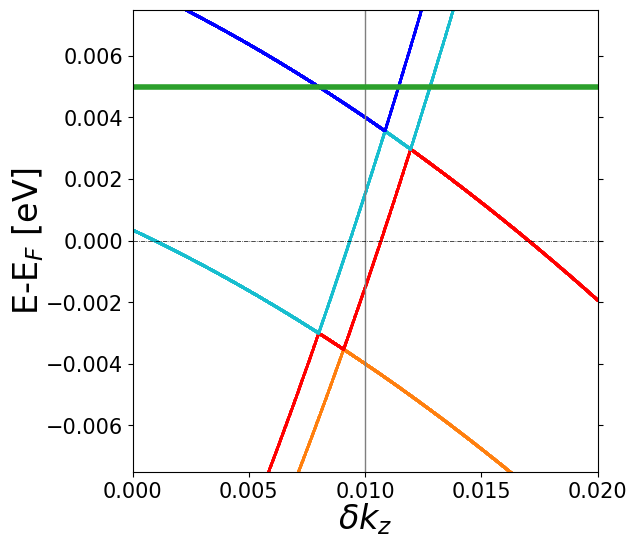}\includegraphics[width=35mm]{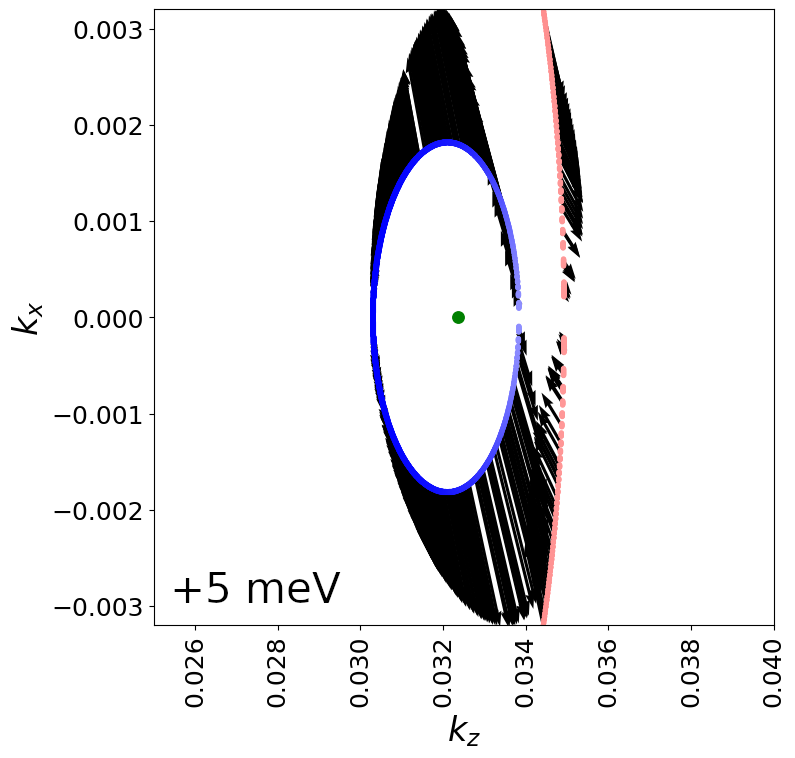}
\hspace{2cm}\includegraphics[width=37mm]{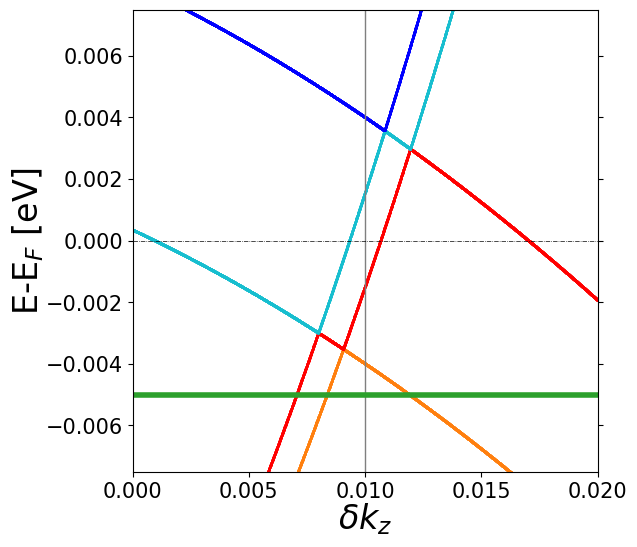}\includegraphics[width=35mm]{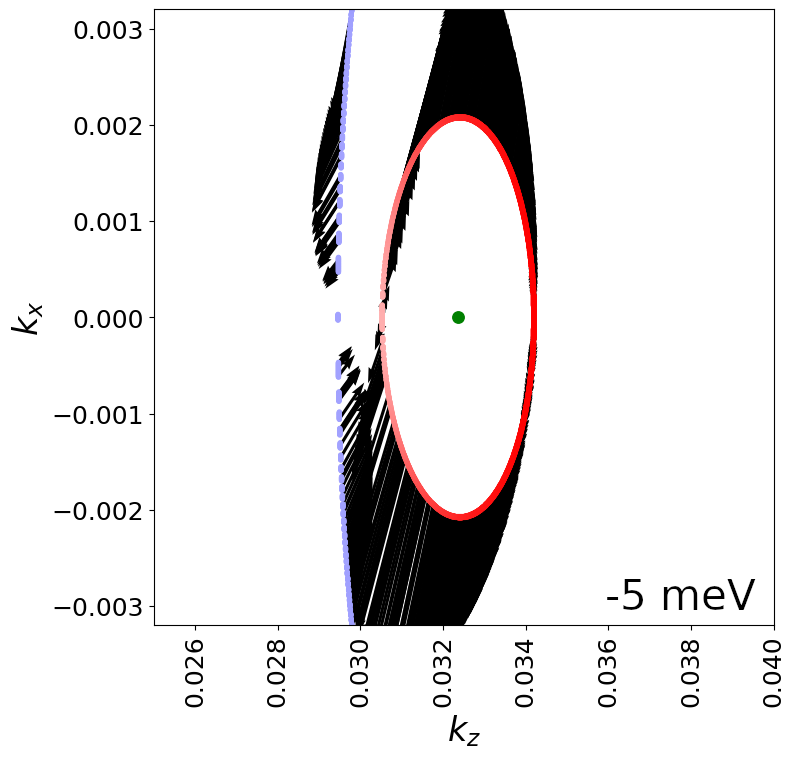}

\includegraphics[width=37mm]{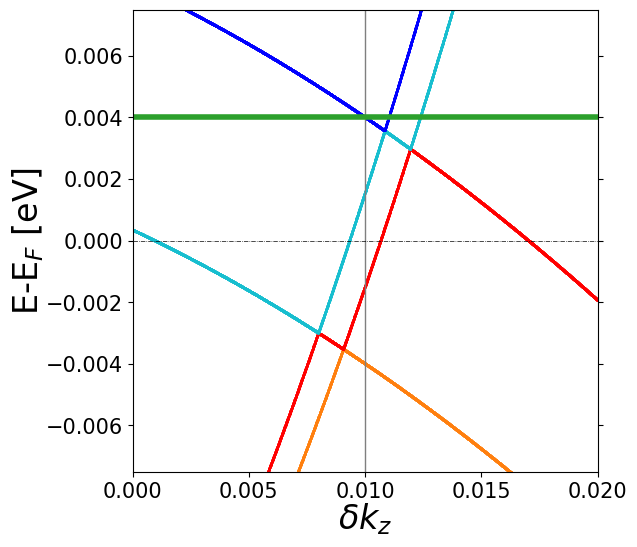}\includegraphics[width=35mm]{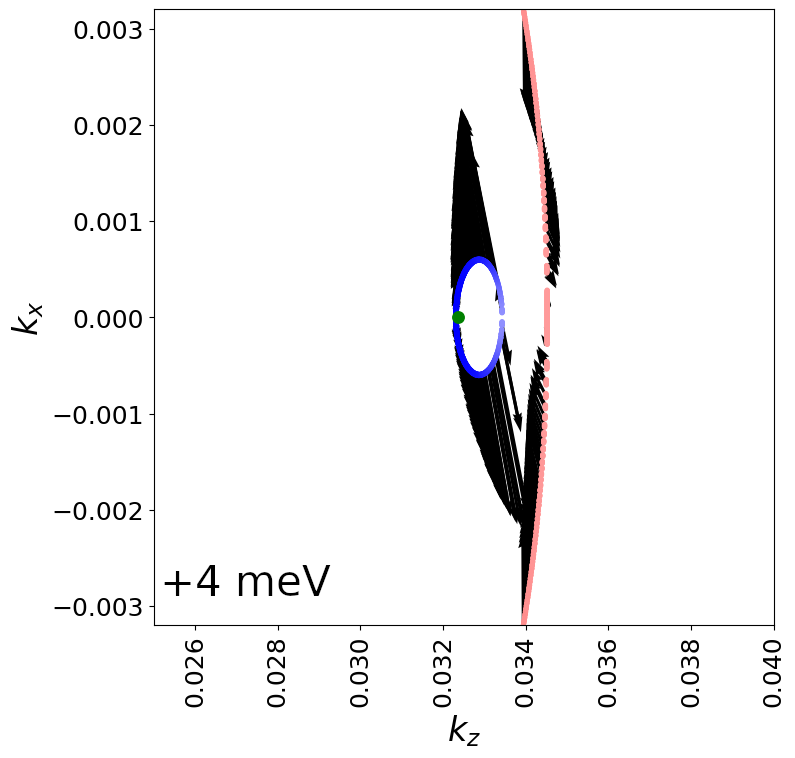}
\hspace{2cm}\includegraphics[width=37mm]{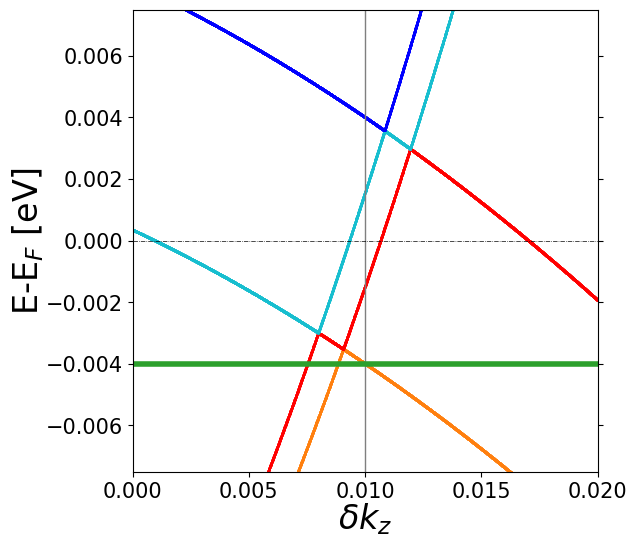}\includegraphics[width=35mm]{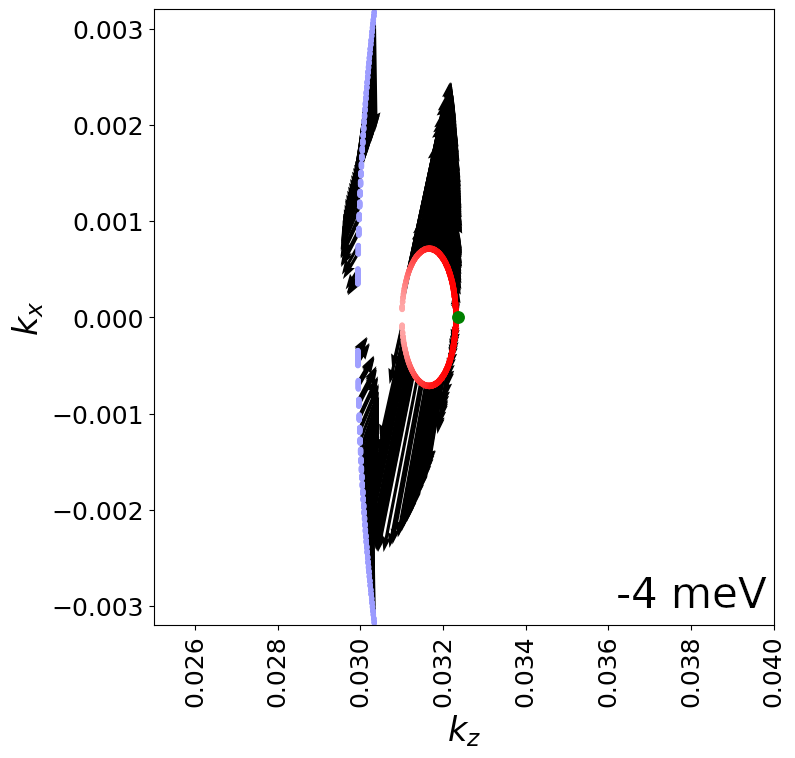}

\includegraphics[width=37mm]{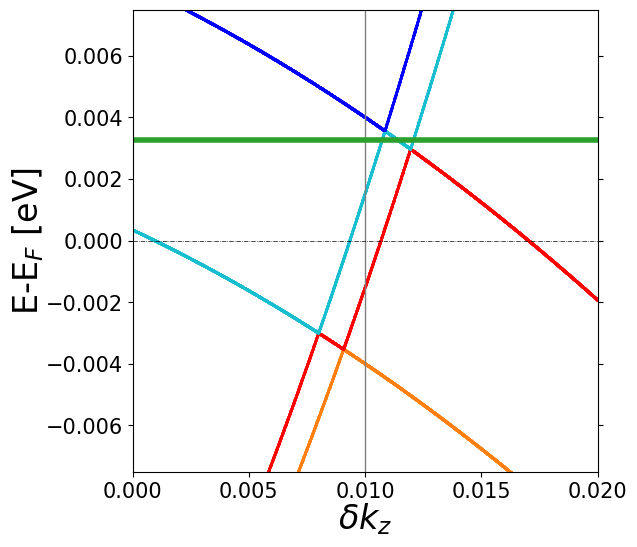}\includegraphics[width=35mm]{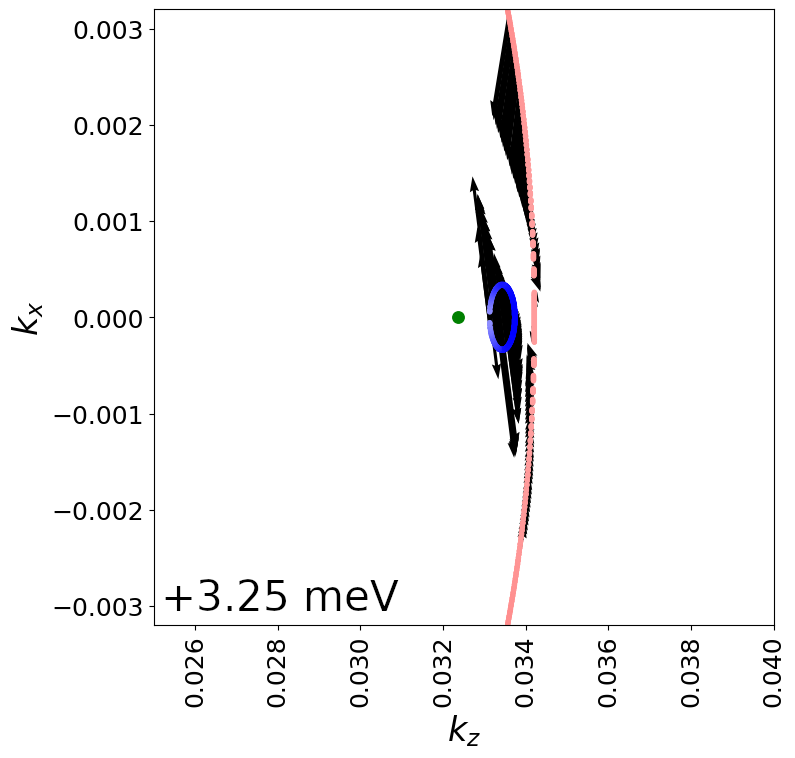}
\hspace{2cm}\includegraphics[width=37mm]{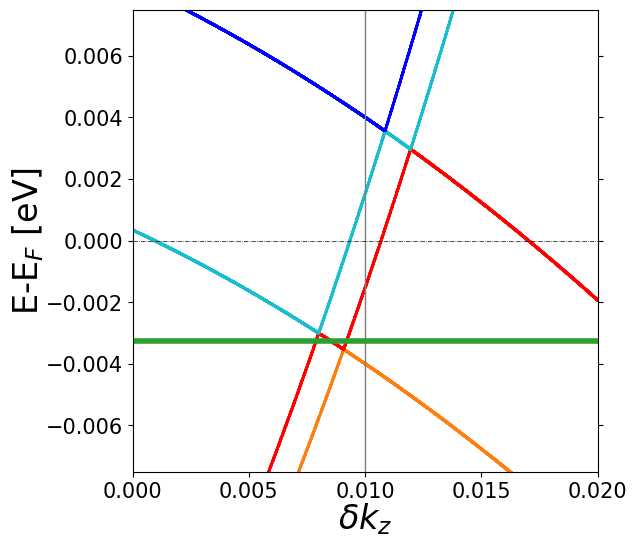}\includegraphics[width=35mm]{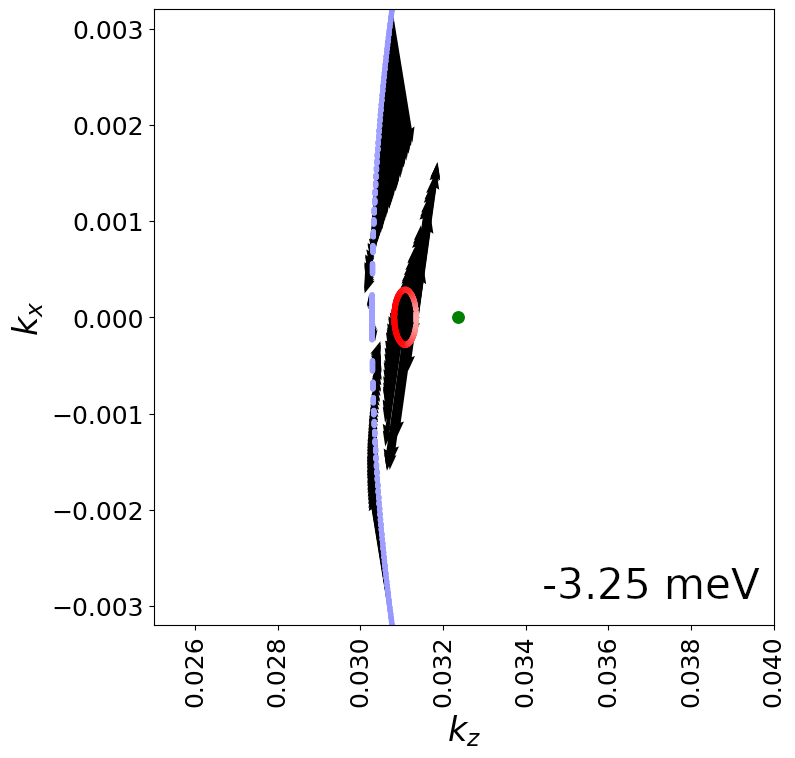}

\includegraphics[width=37mm]{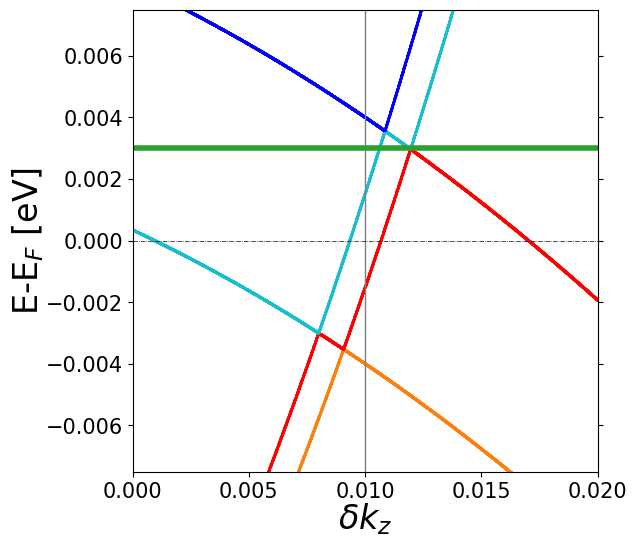}\includegraphics[width=35mm]{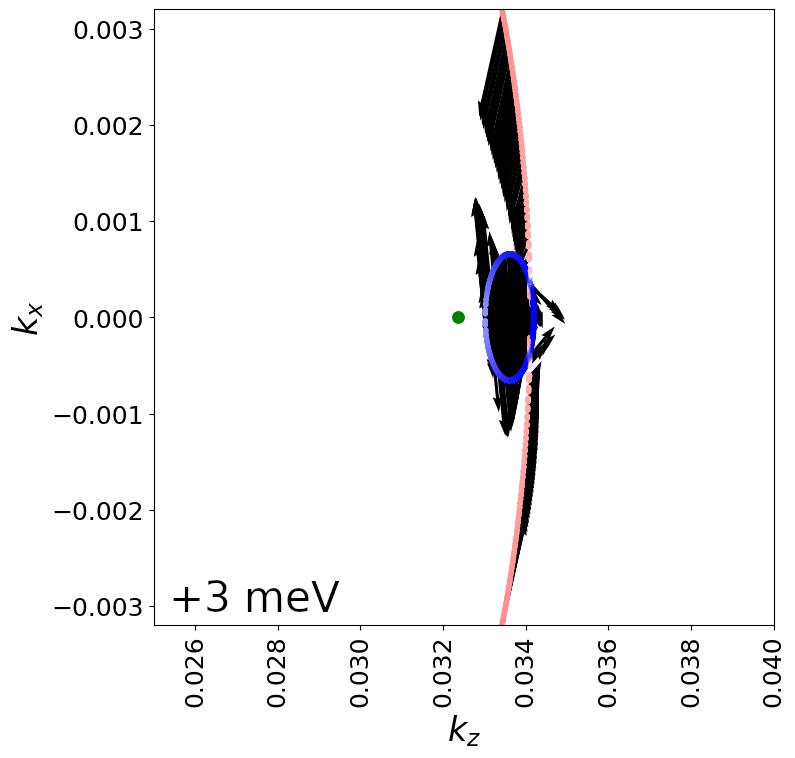}
\hspace{2cm}\includegraphics[width=37mm]{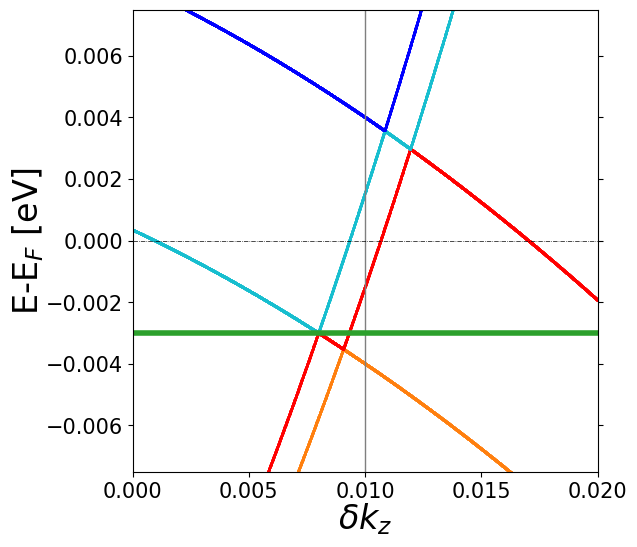}\includegraphics[width=35mm]{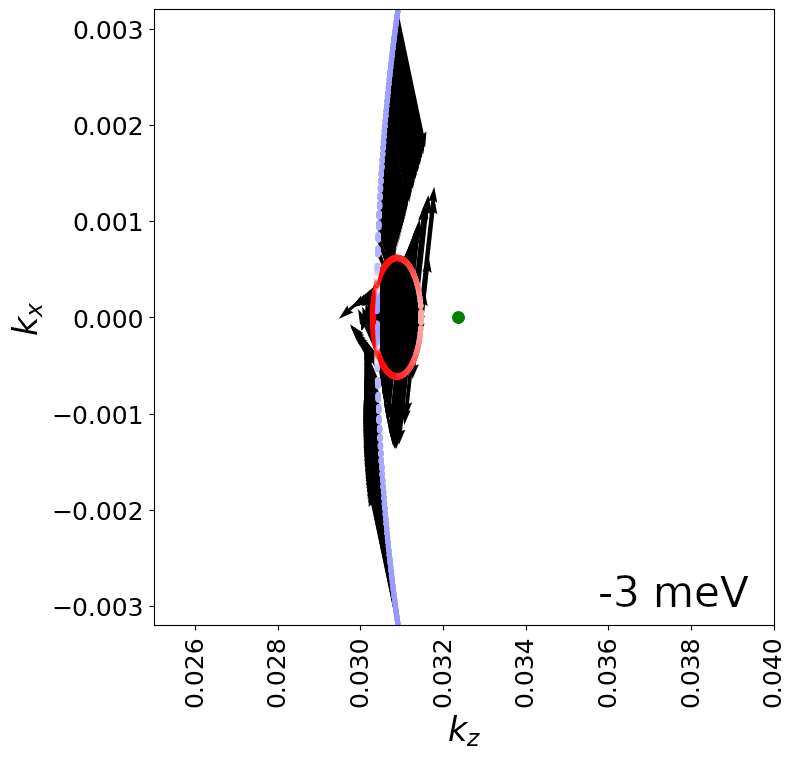}

\includegraphics[width=37mm]{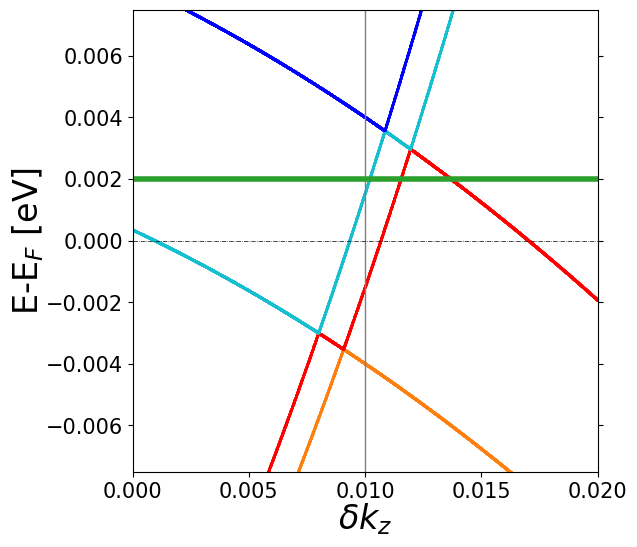}\includegraphics[width=35mm]{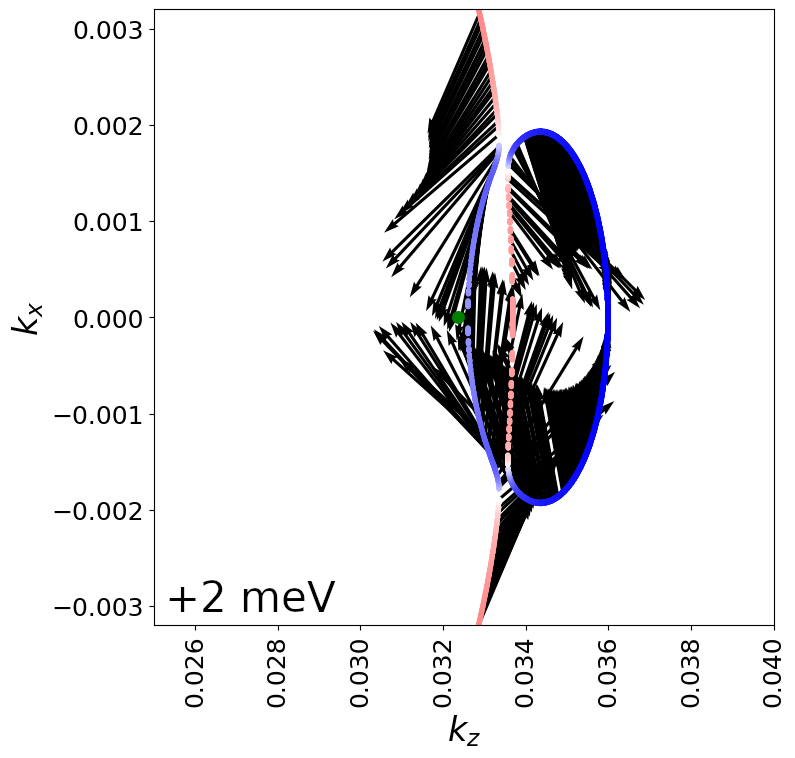}
\hspace{2cm}\includegraphics[width=37mm]{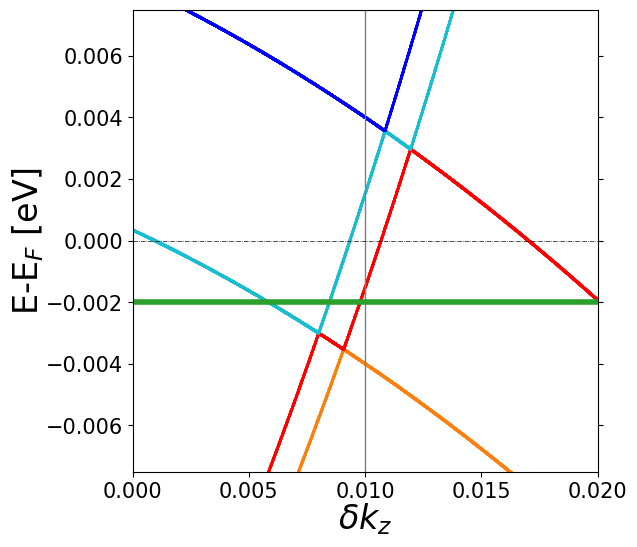}\includegraphics[width=35mm]{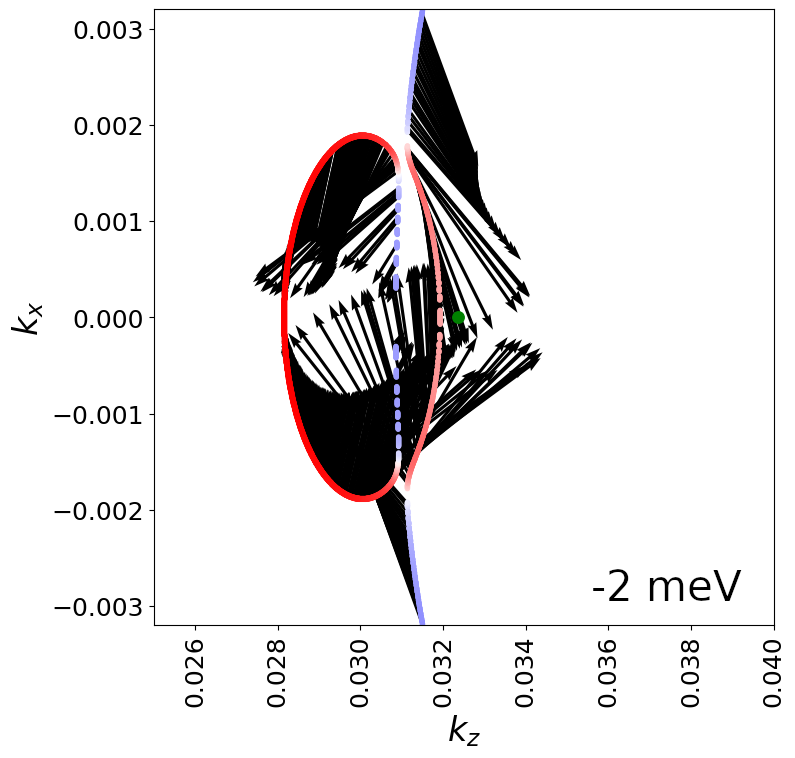}

\includegraphics[width=37mm]{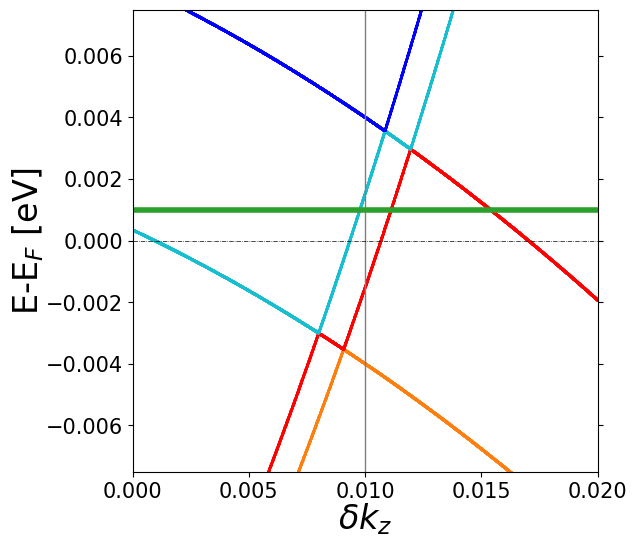}\includegraphics[width=35mm]{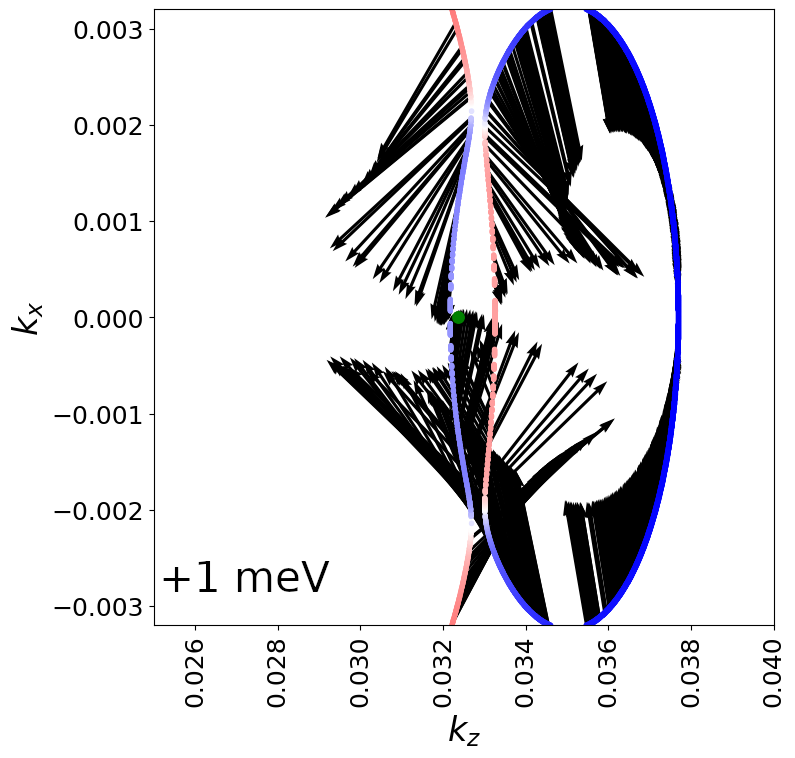}
\hspace{2cm}\includegraphics[width=37mm]{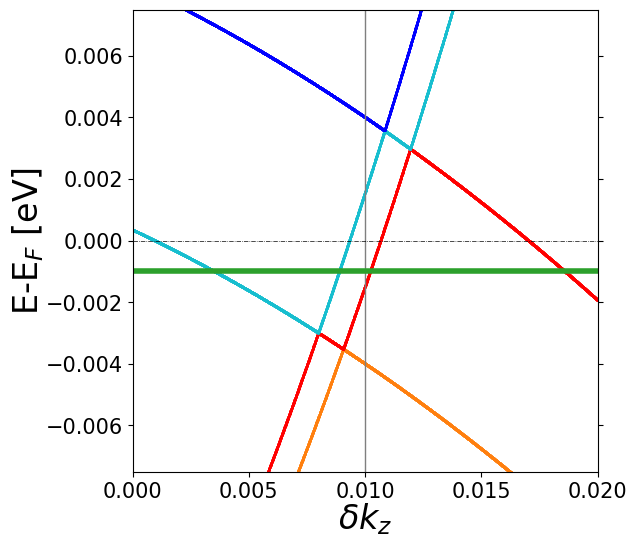}\includegraphics[width=35mm]{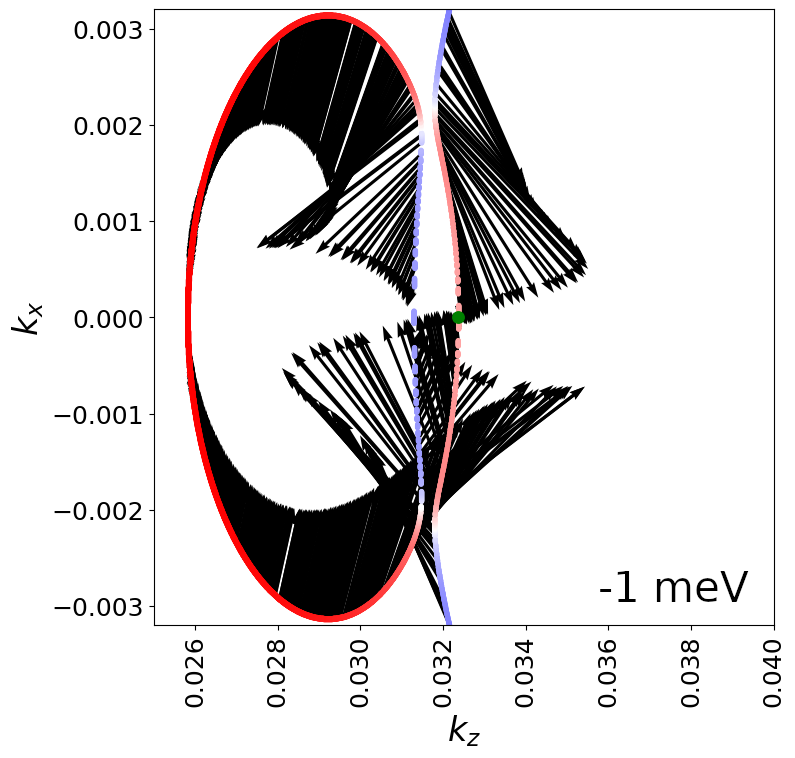}

\includegraphics[width=37mm]{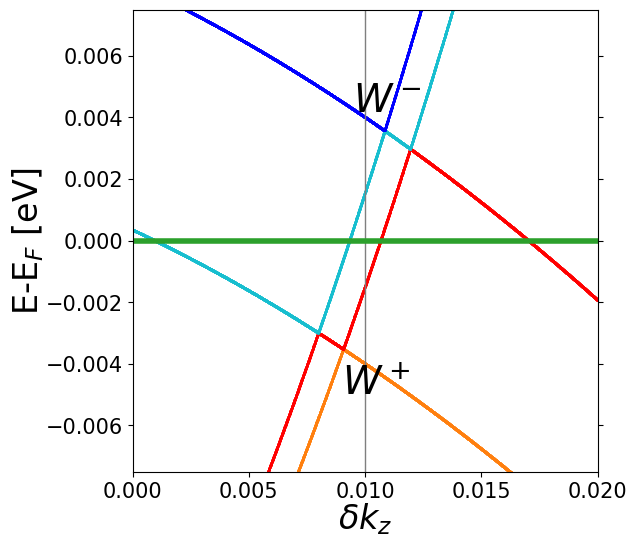}\includegraphics[width=35mm]{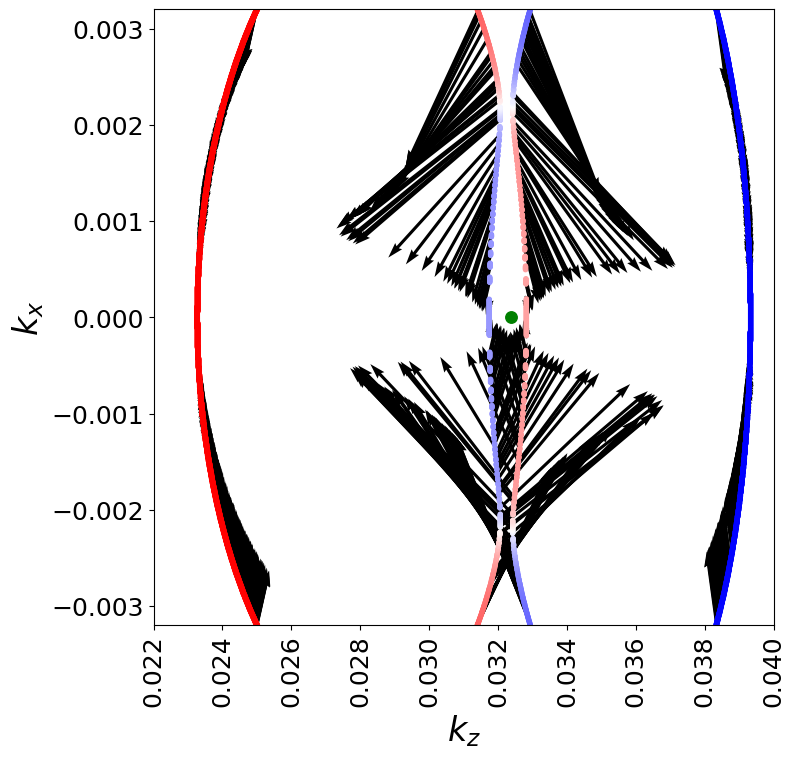}
\caption{\label{fig:app_D2W_bz}
Left and Right double-column panels show the band structure (left side in a double-column)
along $k_z$ path and the 2D maps of constant-energy band structure (right side in a double-column) 
in the $k_z-k_x$ plane around the Dirac point (green dot). The constant energy
is shown by the horizonal green solid-line in the $k_z$ bands.
The applied field is $\gamma B_z$ = 4 meV.
The band coloring in the 2D maps corresponds to the value of $S_z$, i.e. blue $S_z=-1$, red $S_z=+1$,
white $S_z=0$. The black arrows correspond the spin vectors $(S_y,S_x)$ in the $k_z-k_x$ plane.
}
\end{figure}

\newpage

\twocolumngrid 

\begin{figure}
\centering
\includegraphics[width=37mm]{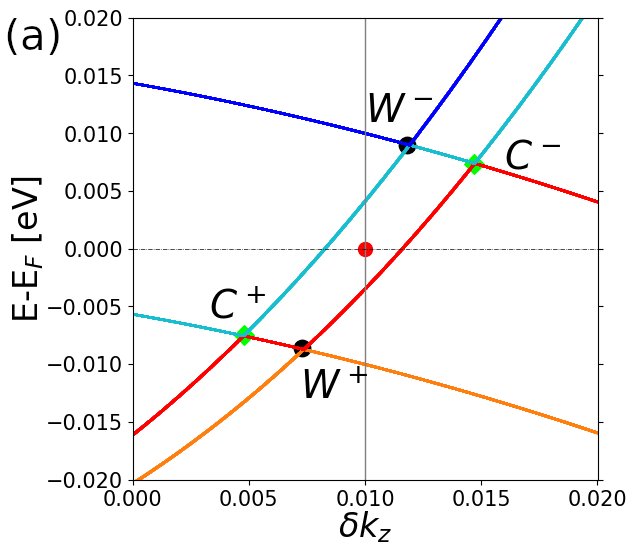}
\includegraphics[width=42mm]{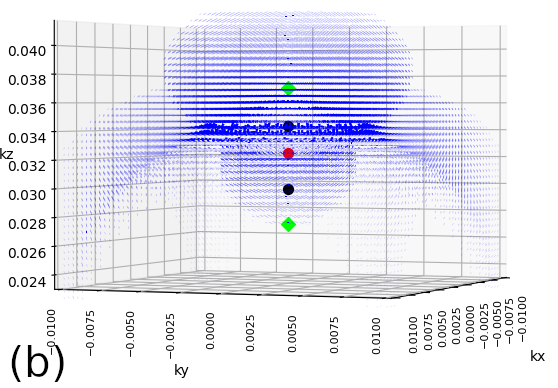}\includegraphics[width=44mm]{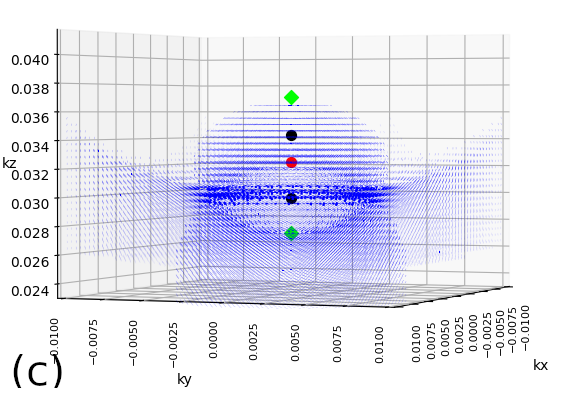}
\caption{\label{fig:app_WandCpts_bz}
{
Panel (a): Band structure with the Weyl points, $W^+$ and $W^-$ (black dots) and  
crossings $C^+$ and $C^-$ (green diamonds) for an applied field is $\gamma B_z$ = 10 meV. 
The red dot represents the original Dirac point.
Panel (b,c): 3D maps, on regular $k$-space grid, of the Berry curvature vectors $(\bar\Omega_x,\bar\Omega_y,\bar\Omega_z)$
(small blue arrows).
The line of symbols, at $k_x=k_y=0$, represent the sequence of $C^+,W^+,D,W^-,C^-$ points along $k_z$, 
with the Dirac point $D$ (red dot) at $k_z = k_D = 0.032\ \AA^{-1}$.
A larger contribution of Berry curvature is found around the Weyl points $W^\pm$ (black dots) than
around the $C^\pm$ crossing points (green diamonds).
Energy window integration for $\bar\Omega_\gamma$ is around the pair of points $W^-,C^-$ (located above the $D$ point, $k_z > k_D$) 
in panel (b); around the pair $C^+,W^+$ (located below the $D$ point, $k_z < k_D$) in panel (c).
}%
}
\end{figure}

\newpage

\end{document}